\documentclass[manuscript]{aastex}
\usepackage{natbib}
\usepackage[caption=false]{subfig}
\usepackage{fancyhdr}
\usepackage{graphicx,color}
\usepackage{ifpdf}
\ifpdf
  \DeclareGraphicsExtensions{.pdf}
\else
  \DeclareGraphicsExtensions{.eps}
\fi
\usepackage{longtable}
\usepackage{tabularx}
\usepackage{threeparttablex}
\usepackage{amsmath}
\usepackage{lscape}
\usepackage{subfig}
\usepackage{amssymb}

\newcommand{\Msun}{\mbox{\,$\rm M_{\odot}$}}
\newcommand{\Lsun}{\mbox{\,$\rm L_{\odot}$}}
 
\newcommand{\teff}{$T_{\rm eff}$}

\newcommand{\logg}{$\log g$}
\newcommand{\mv}{$\xi_{\rm t}$}
\newcommand{\feIh}{$\textrm{[Fe\,I/H]}$}
\newcommand{\feIIh}{$\textrm{[Fe\,II/H]}$}
\newcommand{\feh}{$\textrm{[Fe/H]}$}
\newcommand{\feI}{$\textrm{Fe\,I}$}
\newcommand{\feII}{$\textrm{Fe\,II}$}
 
\shorttitle{A post-AGB star that failed the third dredge-up}
\shortauthors{Kamath et al.}

\begin{document}


\title{Discovery of a metal-poor, luminous post-AGB star that failed
  the third dredge-up.}


\author{D. Kamath \and H. Van Winckel}
\affil{Instituut voor Sterrenkunde, K.U.Leuven, Celestijnenlaan 200D bus 
2401, B-3001 Leuven, Belgium}

\author{P. R. Wood \and M. Asplund}
\affil{Research School of Astronomy and Astrophysics,
Australian National University, Canberra ACT 2611, Australia}

\author{A.I. Karakas}
\affil{Research School of Astronomy and Astrophysics,
Australian National University, Canberra ACT 2611, Australia}
\affil{Monash Centre for Astrophysics, School of Physics and Astronomy,
Monash University, VIC 3800, Australia}

\author{J. C. Lattanzio}
\affil{Monash Centre for Astrophysics, School of Physics and Astronomy,
Monash University, VIC 3800, Australia}




\begin{abstract}

Post-asymptotic giant branch (post-AGB) stars are known to be chemically diverse. In this paper we
present the first observational evidence of a star that has failed the
third dredge-up (TDU). J005252.87-722842.9 is a A-type (\teff\,=\,8250\,$\pm$\,250K) luminous (8200\,$\pm$\,700\,\Lsun), metal-poor
(\feh\,=\,$-1.18\,\pm$\,0.10), low-mass (M$_{\rm initial}$\,$\approx$\,1.5\,$-$\,2.0\,\Msun) post-AGB star in the Small Magellanic
Cloud. Through a systematic abundance study, 
using high-resolution optical spectra from UVES, we found that this
likely post-AGB object shows an intriguing photospheric composition
with no confirmed carbon-enhancement (upper limit of
  [C/Fe]\,$<$\,0.50) nor
enrichment of $s$-process elements. We derived an oxygen abundance of
[O/Fe]\,=\,0.29\,$\pm$\,0.1. For Fe and O, we
took into account the effects of non-local thermodynamic equilibrium
(NLTE). We could not derive an upper limit
for the nitrogen abundance as there are no useful nitrogen lines within our spectral coverage. 
The chemical pattern displayed by this object
has not been observed in single or binary post-AGBs. Based on its derived stellar parameters and inferred 
evolutionary state, single star nucleosynthesis
models predict that this star should have undergone TDU episodes
while on the AGB and be carbon-enriched. However, our
observations are in contrast with these
predictions. We identify two possible Galactic analogues which are likely to be post-AGB stars, but the lack of
accurate distances (hence luminosities) to these objects does not
allow us to confirm their post-AGB status. If they have low luminosities then they are likely to be dusty post-RGB stars. The discovery of
J005252.87-722842.9 reveals a new stellar evolutionary
channel whereby a star evolves without any third dredge-up episodes. 

\end{abstract}


\keywords{(galaxies:) Magellanic Clouds, methods: observational, stars: abundances, stars: AGB and
  post-AGB, stars: chemically peculiar, stars: evolution}



\section{Introduction}
\label{intro}

Post-asymptotic giant branch (post-AGB) stars are low- to intermediate-mass stars that have evolved
off the AGB because a strong dusty mass loss removed most of the 
stellar envelope. This mass loss occurs either by binary interaction
or, for single stars, by a phase of very high 
mass-loss called the superwind \citep[e.g.,][and references therein]{nie12}. Their
atmospheres contain signatures of the chemical enrichment from internal nucleosynthesis that
has occurred prior to and during their entire AGB
lifetime. For single stars, during the thermally pulsing AGB (TP-AGB)
phase, the photospheric abundance of the
object can be altered in two ways. Firstly, via third dredge-up (TDU)
episodes that can occur after a thermal pulse. These are responsible for
enriching the stellar photosphere with products of internal
nucleosynthesis such as carbon (C), nitrogen (N), oxygen (O) 
and heavy elements produced by the slow neutron capture process
\citep[$s$-process, see][for a review]{karakas14a}. They can also convert stars into carbon-rich (C-rich) stars with a C/O
ratio greater than unity \citep{herwig05}. Secondly, via hot bottom burning (HBB),
which is thought to be only active in stars of initial mass greater than 4\,$-$\,5\,\Msun. HBB
prevents the formation of carbon-rich photospheres by burning $^{\rm
  12}$C into $^{\rm 14}$N
\citep{boothroyd93,lattanzio92,lattanzio96,karakas07b,ventura15}. We note that stellar
evolutionary sequences and nucleosynthesis models are metallicity and
mass dependant and are subjected to many uncertainties
\citep[see][for a review]{karakas14a}.

During the post-AGB phase, the warm
stellar photosphere makes it possible to quantify photospheric
abundances for a very wide range of elements from
CNO up to some of the heaviest $s$-process elements well beyond the 
Ba peak \citep{reyniers03}, that are brought to the 
stellar surface during the AGB phase \citep{busso01,cristallo11,fishlock14}.
Therefore, post-AGB objects can provide direct and stringent
constraints on the parameters governing stellar evolution and
nucleosynthesis, especially during the chemically-rich AGB phase.

Post-AGB stars are surrounded by the relic of a heavy mass-loss
episode. Therefore, they have a dusty circumstellar environment which
is characterised by a large mid-infrared (mid-IR) excess. This has allowed for the identification and study of
post-AGB stars in the Galaxy and in the Magellanic Clouds. 

Extensive studies using Galactic post-AGB objects have also shown that the
majority of the optically visible post-AGB stars can be
classified into two groups based on their spectral energy
distributions \citep[SEDs, see][for a review]{vanwinckel03}. The first group of objects, referred to as shell-type sources, display a double
 peaked SED. The peak at shorter wavelengths is due to emission from the stellar photosphere and the peak at longer wavelengths is due to circumstellar dust which usually peaks at wavelengths greater than
 10$\mu$m. This has been illustrated from radiative transfer models of a well known
 expanding shell source HD161796, where the peak of the dust SED is at
 around 30$\mu$m \citep{min13}. This type of SED is most
 likely characteristic of single post-AGB stars surrounded by an 
 optically thin expanding circumstellar envelope that is the remnant
 AGB mass-loss. 
The shell continues to move outwards, gradually exposing the central
star, resulting in a double-peaked SED. These objects mostly contain
the photospheric chemistry that is representative of single stars.
The second group of
objects display SEDs with a clear near-infrared ($\lambda$\,$<$\,10\,$\mu$m) excess indicating
  that circumstellar dust must be close to the central star, near the 
sublimation temperature. It is now 
well established that
    this feature in the SED indicates the presence of a stable compact
    circumbinary disc (see Van Winckel 2003, for a
review). The commonly accepted evolutionary scenario that 
results in the formation of a circumbinary disc is that during the
    star's ascent up the AGB, it suffers a strong interaction with a
    companion (likely to be an unevolved main-sequence star). 
The AGB star fills its Roche lobe and stable mass transfer ensues
which results in the formation of a circumbinary disc. What we observe
is then a F- or G-type post-AGB supergiant in a binary system, surrounded by a
circumbinary dusty disc. Therefore dusty discs are considered to be characteristic of binary post-AGB
    stars. These sources are referred to as
    disc-sources
    \citep[e.g.,][]{deruyter06,deroo07,gielen11,hillen13}. The
    rotation of the disc has been resolved with the ALMA array \citep{bujarrabal13a,bujarrabal15} in
  two objects and our survey using single dish observations
  confirms that disc rotation is indeed widespread
  \citep{bujarrabal13b} .

Galactic post-AGB objects not only show a diversity in SED
characteristics but are also chemically more diverse than
anticipated. \citet{kwok89} found an emission feature at 21 microns
that was only observed in C-rich post-AGB stars (thereafter known as 21\,micron
sources), with a varying strength but constant line profile
\citep{volk99}. \citet{klochkova95} discovered strong $s$-process
lines in the Galactic 21\,micron source HD56126. Subsequently, as
expected for these objects, many studies \citep[e.g.,][]{vanwinckel00,reddy02,reyniers03,reyniers07} revealed $s$-process overabundances in
objects with clear C-rich dust. These objects are likely to have
evolved via the single star
evolutionary scenario. It was also found that post-AGB objects that
have O-rich dust chemistry are not $s$-process enriched
\citep{vanwinckel03}. An example is the dusty high-latitude A-F supergiants, which are
metal poor, C-enhanced, more strongly N-enhanced, and not enhanced or deficient in
$s$-process elements \citep{luck90}. Furthermore, many non $s$-process enriched objects show a
characteristic chemical pattern of 'depletion' of refractory elements
in their photosphere whereby elements with a high dust-condensation
temperature (such as Fe, Ti, Ca and Cr) are systematically under-abundant
\citep{vanwinckel98a,giridhar00,maas05,deroo05a,gezer15}.  The process to acquire this chemical anomaly is not completely 
understood, but it is believed that radiation pressure on the
circumstellar dust grains results in a chemical fractionation of dust
and gas in the circumstellar environment. The cleaned gas (with C, N,
O, S and Zn) is then 
re-accreted onto the stellar surface making the photosphere devoid of refractory
elements \citep[see review by][]{vanwinckel03}. 
As a result of this process, the stars display a peculiar photospheric
composition similar to the interstellar gas. \citet{waters92}  proposed that the most likely 
circumstance for the process to occur is when the dust is trapped in
  a circumstellar disc. In almost all depleted post-AGB objects, there
  is observational evidence that a stable circumbinary disc is present
  \citep{deruyter06,gezer15} which seems to be a prerequisite to obtain photospheric
  depletion patterns. Therefore it is likely that these objects
  follow the binary evolution scenario mentioned above. Furthermore,
  the circumstellar dust in the disc is O-rich \citep{gielen08} implying
  that formation of the disc occurred when the object was still an
  M-type AGB star. In some disc sources double chemistry is detected in
  which the C-rich component is mainly limited to polycyclic aromatic
  hydrocarbon (PAH) emission
  \citep{gielen09}. This does not necessarily mean that the central
  object is a carbon star, as PAH emission is also found in
  environments where dissociation of CO can liberate C atoms
  \citep{geballe89,witt09,guzman11}. 

Though the Galactic sample is observationally well studied and has revealed
important chemical and morphological characteristics of post-AGB
stars, the so-far poorly constrained distances and hence initial masses and
luminosities, pose a severe limitation on our ability to fully exploit
this assorted group of objects. The recent first Gaia data
  release, \citep[Gaia DR1,][]{gaia16}, has provided parallaxes to a
  small sample of known Galactic post-AGB stars however the errors in their parallaxes are too
  large to be able to accurately determine their distances. Unknown distances (and
hence luminosities and initial masses) hamper the interpretation of
the diversity in the photospheric chemistry as well as the morphology of
these objects as a function of luminosity and mass.

In our recent studies, we  exploited the release of the infrared
Large Magellanic Cloud (LMC) and Small Magellanic Cloud (SMC) SAGE-Spitzer
  surveys of \citet{meixner06} and \citet{bolatto07} to identify optically bright evolved stars
  with infrared colours indicative of a past history of heavy dusty
  mass loss in the LMC \citep{vanaarle11,kamath15} and SMC
  \citep{kamath14}. Realising the importance of the
  spectral characterisation of the central objects to confirm their
  post-AGB nature, we performed a low-resolution optical spectral
  survey with the AAOmega multi-fibre spectrograph with a 2 degree
  field of view \citep{lewis02}, mounted on the Anglo Australian 3.9m telescope.  Our detailed spectral analysis resulted in a clean
    and extensive census of well characterised post-AGB stars with
    spectroscopically determined stellar parameters T$_{\rm eff}$,
    \,$\log\,g$, \feh\, and E($B-V$) spanning a wide range of 
    luminosities in the LMC \citep{kamath15} and SMC
    \citep{kamath14}. Additionally, our survey also resulted in the
    discovery of dusty post-red giant branch stars (post-RGB) stars, which display similar stellar parameters and SED
    characteristics as post-AGB stars but have lower luminosities
    (L/\Lsun\,$\leq\,$2500). These objects are believed to have
    evolved off the RGB via binary interaction \citep[see][]{kamath16}.

By virtue of their spectral types, favourable bolometric
corrections and especially their constrained distances, the Magellanic
Cloud (MC) post-AGB stars 
offer unprecedented tests for AGB theoretical structure and 
enrichment models of low- and intermediate-mass stars. 
The photospheric chemistry of a small fraction of these post-AGB stars in
the MCs have been studied in detail until now. The first chemical abundance
studies, carried out by \citet{reyniers07} and
\citet{gielen09}, showed a depletion pattern. As expected, all these
objects had dusty circumstellar discs, which, together with their
depleted photospheric chemistry points to their binary nature.  Our 
recent high-resolution chemical abundance studies, also 
resulted in the discovery of the most $s$-process enriched post-AGB star in the SMC
\citep{desmedt12}. This
object is likely to be a single, luminous (L/\Lsun\,$\approx\,$7000), C-rich, 21\,micron
source with shell-type SED. 
In our subsequent studies, we found more $s$-process enriched objects
with shell-type SEDs \citep{vanaarle13,desmedt15}. The MC post-AGB
objects therefore also follow 
the same chemical diversity and morphology as that of the Galactic
objects. 

In this paper we present a detailed chemical abundance study of a 
chemically peculiar post-AGB star, \objectname{J005252.87-722842.9} 
(hereafter referred to as J005252), that resides in the SMC. In the following sections of
the paper we present our detailed spectral analysis (determination of
stellar parameters as well as the abundance analysis) of
\objectname[J005252.87-722842.9]{J005252} along with an interpretation
of our results.  We also present the likely Galactic analogues of
\objectname[J005252.87-722842.9]{J005252}. Finally, we conclude with a
summary of our study and its implications for stellar evolution. 

\section{Data and Observations}
\label{DO}


\subsection{Photometry}
\label{phot}

\begin{figure}
\begin{center}
\epsscale{1.10}
\plotone{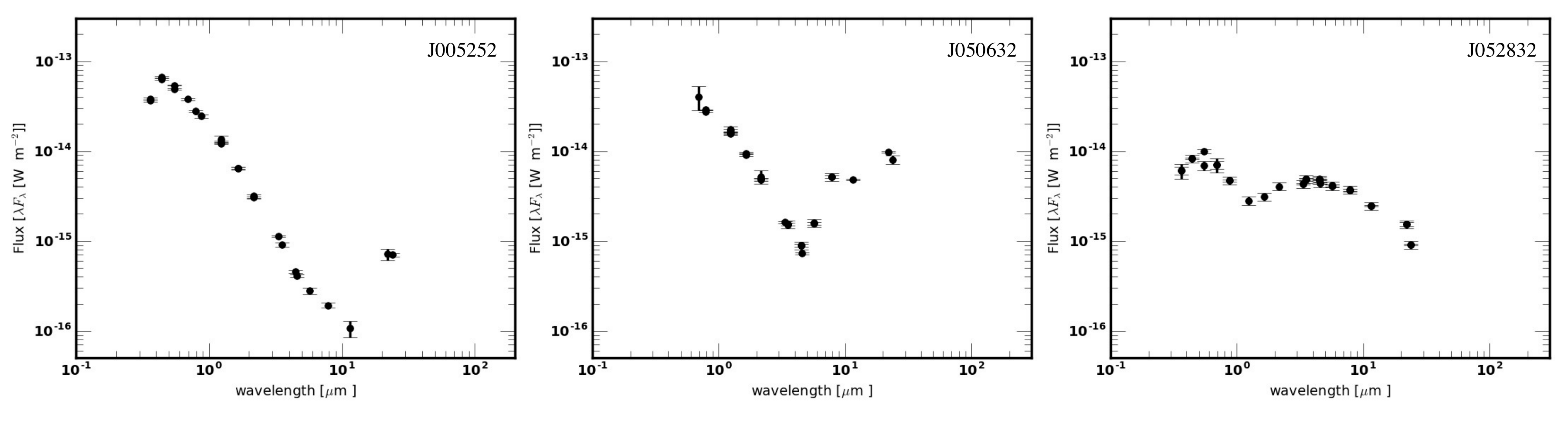}
\caption{Left panel: SED of J005252. Center
  panel: SED of J050632.10-714229.8, a LMC shell-type source \citep{vanaarle13}. Right panel: SED of
  J052832.60-690440.6, a LMC disc-type source \citep{kamath15}. The black symbols
  indicate the broadband photometry corrected for foreground
  extinction.\label{J005252_sed}}
\end{center}
\end{figure}

\objectname[J005252.87-722842.9]{J005252}
was first identified as a post-AGB candidate in our extensive 
low-resolution spectroscopic survey \citep{kamath14}. 
The photometry covering the optical, near-IR and
mid-IR bands are presented in
\citet{kamath14}. Figure~\ref{J005252_sed} shows the SED for \objectname[J005252.87-722842.9]{J005252}. The
 SED is constructed using the broadband photometry. The full procedure
 that we used to correct for foreground
extinction (which takes into account the extinction produced by the interstellar dust in our Galaxy, along the line of
sight of the SMC and from the SMC itself), is mentioned in
\citet{kamath14,kamath15}. In brief, we use the extinction law of
\citet{cardelli89} with a mean reddening, for the combined SMC and
Galactic components, of
E($B$-$V$)\,=\,0.12 \citep{keller06}.

Based on the visual
inspection of the dust excess in the SED of
this object (see Figure~\ref{J005252_sed}), it appears to be 
a shell-type source with an expanding dusty circumstellar
shell. In Figure~\ref{J005252_sed}, we show the SED of a post-AGB
star (J050632.10-714229.8) with a shell-type SED \citep{vanaarle13}
and a post-AGB star (J052832.60-690440.6) with a disc-type SED
\citep{kamath15}. From this comparison, it can be clearly seen that
the SED of \objectname[J005252.87-722842.9]{J005252} is shell-type. The observed luminosity of this object ($L_{\rm obs}$) is
8000\,\Lsun\, \citep[see][]{kamath14}, and was obtained by integrating the SED that
is defined by the photometry corrected for foreground extinction while
applying a distance modulus for the SMC of 18.93 
\citep{keller06}.  The integration 
was not extended beyond the available wavelength points in the flux
distribution. Therefore, the estimated observed luminosity is a lower
limit. The circumstellar extinction was not corrected for because it was assumed that all radiation absorbed by the 
circumstellar matter was re-radiated at a longer mid-IR wavelength still within the wavelength 
range of the observed SED. In Section~\ref{lum+ini_mass} we
estimate the total line-of-sight extinction towards \objectname[J005252.87-722842.9]{J005252}, which is the sum of the
Galactic, SMC and circumstellar reddening. Using this total extinction
we derive the photospheric luminosity by integrating under the full
model atmosphere appropriate for
\objectname[J005252.87-722842.9]{J005252} (see Section~\ref{lum+ini_mass})
which provides a more accurate luminosity of the central star of this
object than the observed luminosity ($L_{\rm obs}$/\Lsun).


\subsection{Spectroscopic observations and data reduction}
\label{spec}

We obtained high-resolution optical spectra using the UVES echelle 
spectrograph \citep{dekker00}, mounted on the 8m UT2 Kueyen Telescope
of the VLT array at the Paranal Observatory of ESO in
Chile. We used the dichroic beam-splitter 
which resulted in a wavelength coverage in the blue arm from
approximately 3280 to 4530\,\AA\, and in the red arm for the lower 
and upper part of the mosaic CCD chip from approximately 4780 to 
5770\,\AA\, and from 5800 to 6810\,\AA, respectively. We chose a slit
width of 1 arcsecond to obtain a good compromise between 
spectral resolution and slit-loss minimalisation. The details of the 
object along with its coordinates are listed in 
Table~\ref{table:fields}.

\begin{figure}
\begin{center}
\epsscale{1.1}
\plottwo{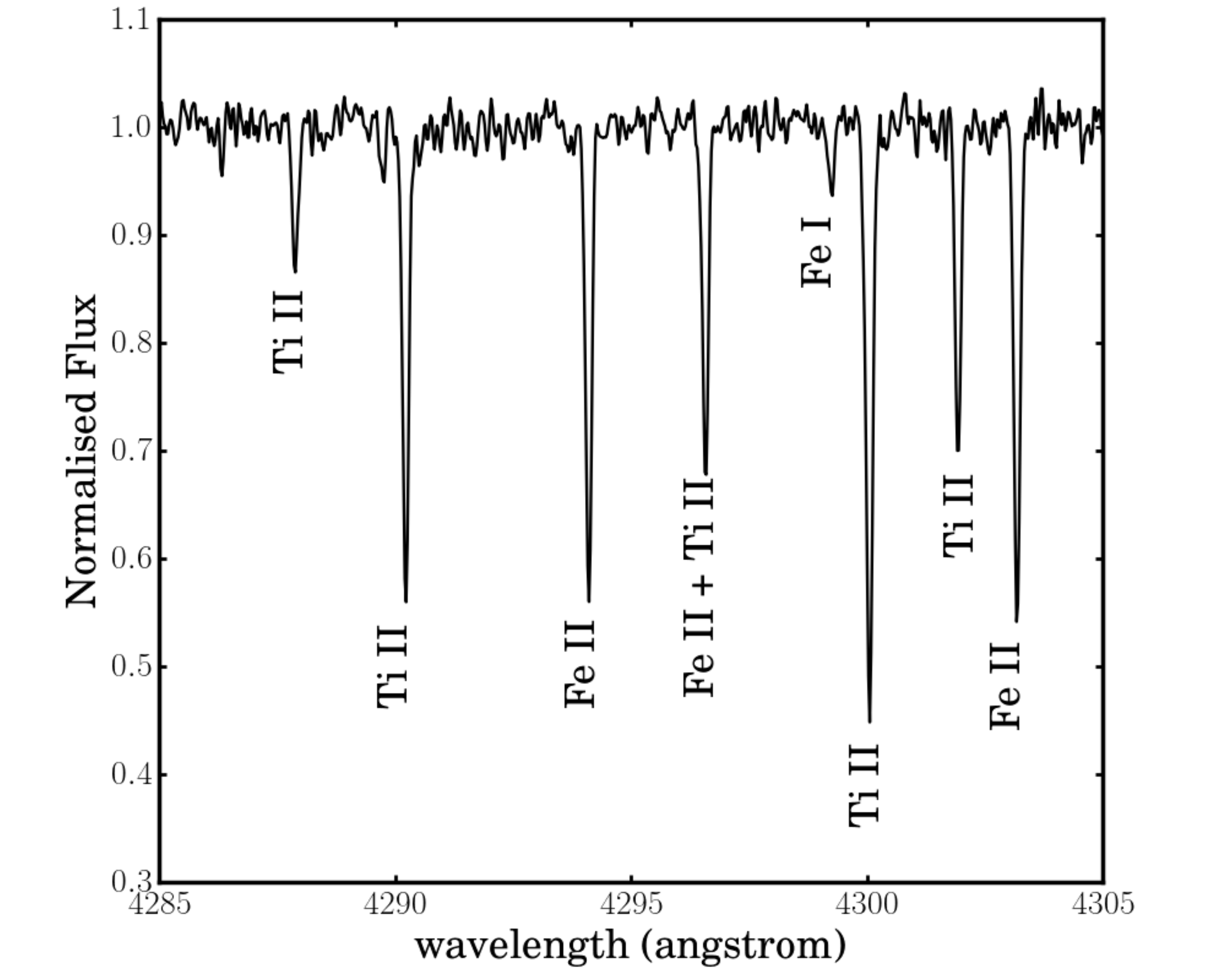}{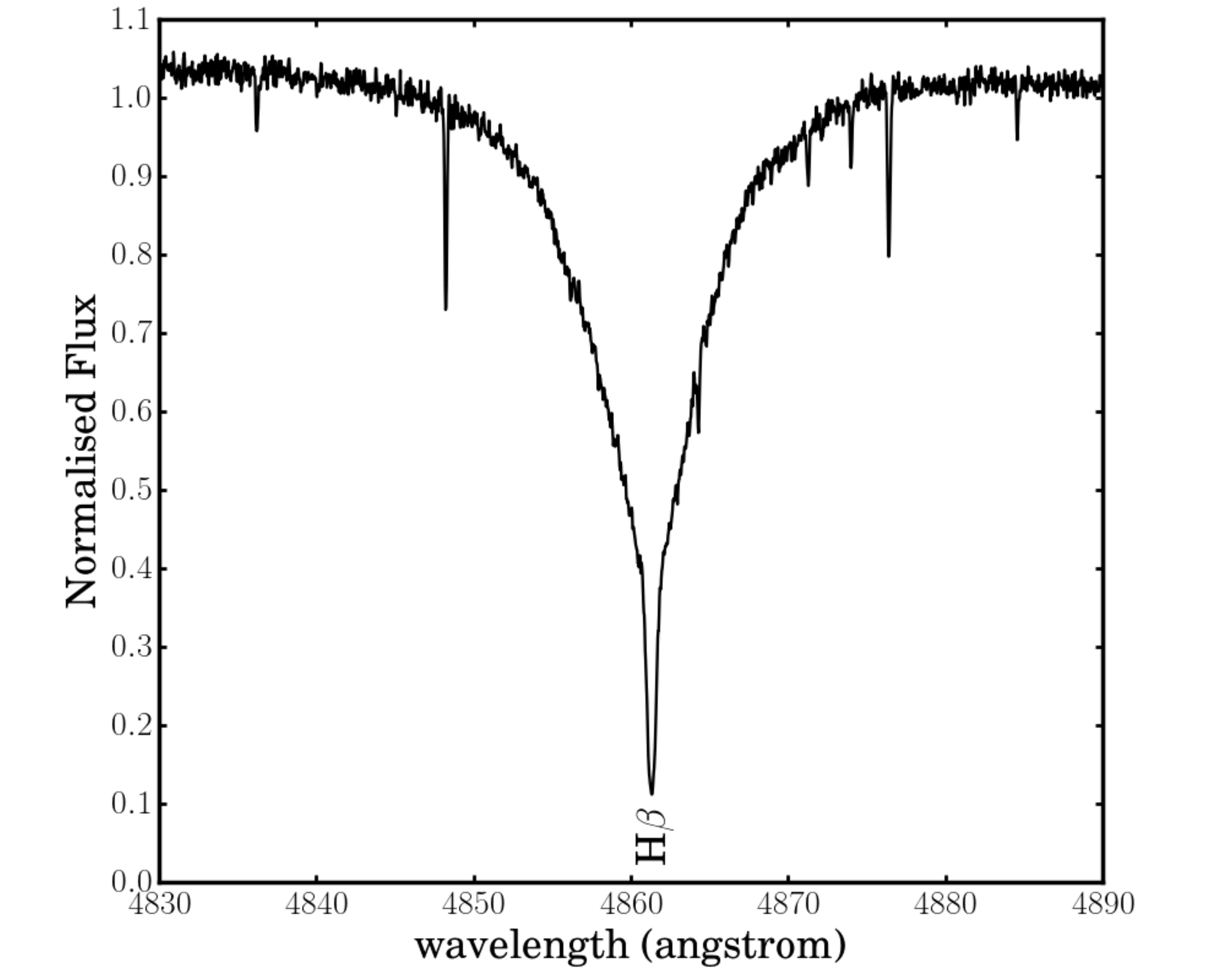}
\plottwo{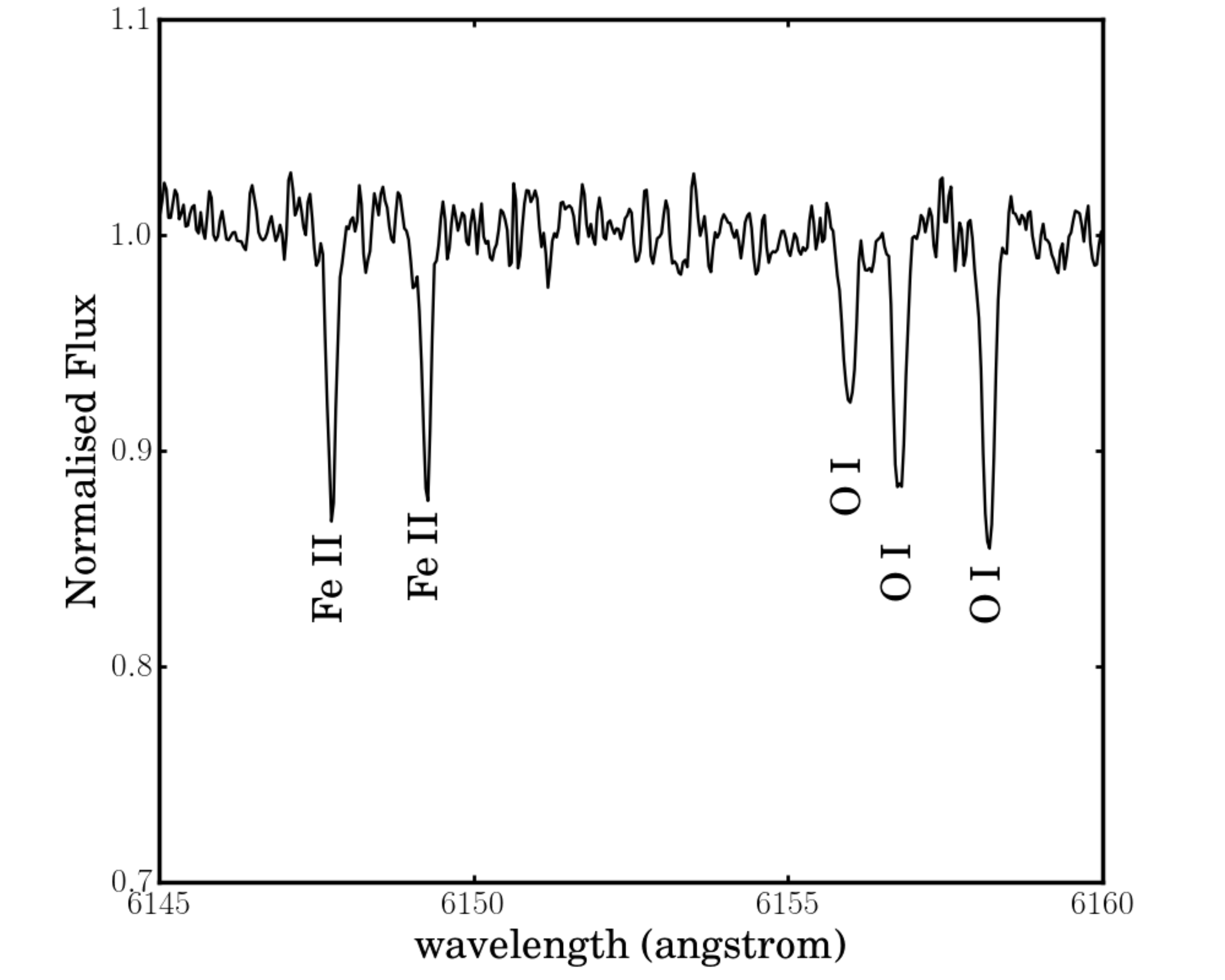}{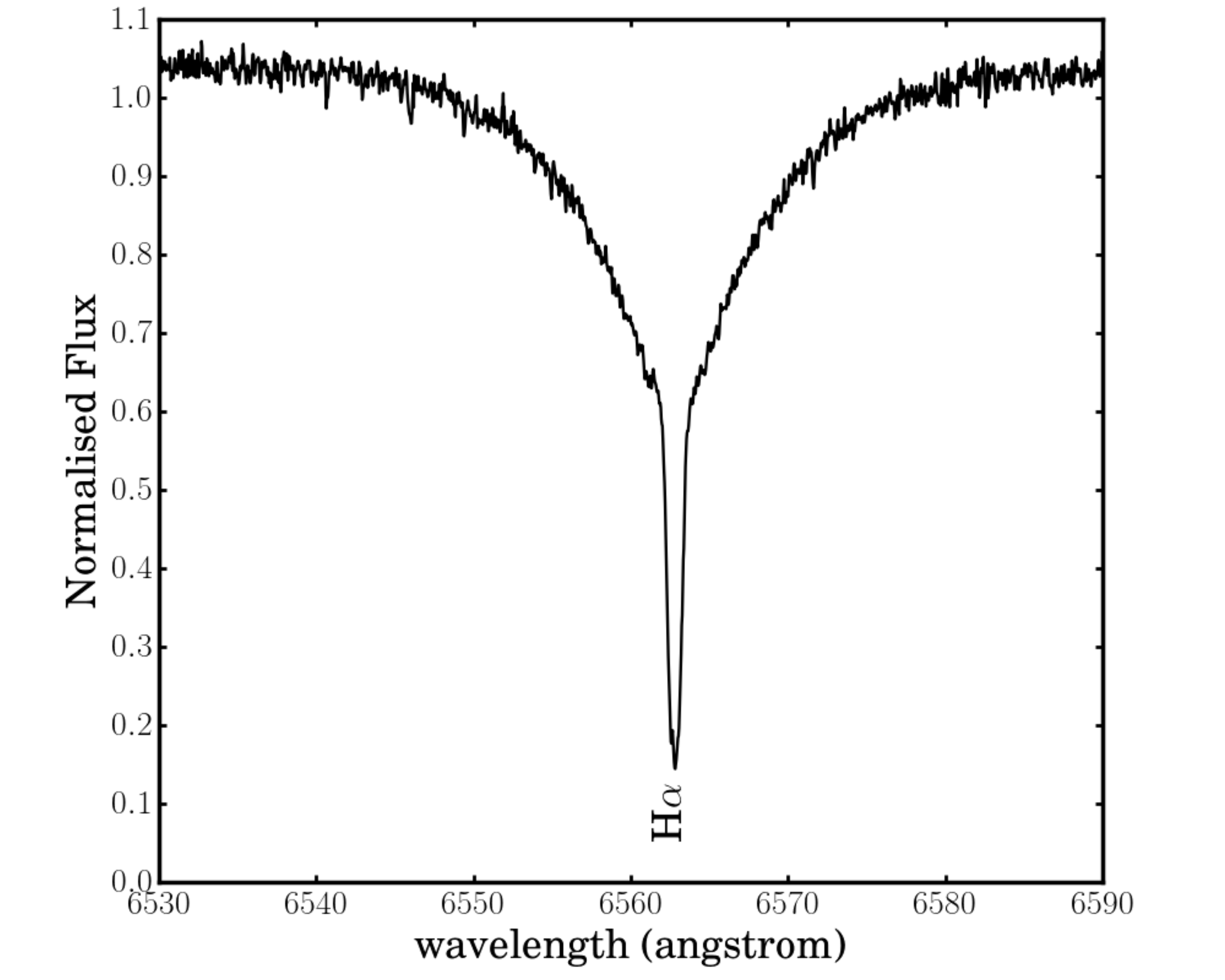}
\caption{Spectral samples of J005252. The top-left panel shows a
  spectral region in the blue part of the spectra with lines of
  titanium (Ti) and iron (Fe). The top-right panel shows the H$\beta$
  line at 4861.36\,\AA. The bottom-left
  panel shows the spectral region that contains neutral oxygen lines,
  the O\,I triplet, at 6155.97, 6156.78 and 6158.19\,\AA. The bottom-right panel shows the H$\alpha$
  line at 6562.80\,\AA. \label{J005252_sample_spectra}}
\end{center}
\end{figure}

The UVES raw data
were reduced using the UVES pipeline (version 5.3.0) in the Reflex
environment of ESO. The reduction process included the standard procedures for echelle data reduction: background correction, extraction of the spectral orders with cosmic-clipping, flat-fielding and wavelength calibration. For spectra reduced within the Reflex 
environment, the standard reduction parameters of the UVES pipeline 
were used as these resulted in the best signal-to-noise (S/N). 
Once reduced, the weighted mean spectra were created for the 
three different wavelength regions. Each sub-spectrum was then
normalised individually by fitting a fifth order polynomial through
interactively defined continuum points. Finally, the subspectra were
merged into one large spectrum, which is used for the spectral
analysis. During merging, the mean flux was calculated for the 
overlapping spectral regions of the subspectra. The final spectrum has
a mean S/N of 90 in the red part and a S/N ranging from 25 to 45 in
the blue (above 4000\,\AA). In
Figure~\ref{J005252_sample_spectra} we show samples of our available spectra to illustrate the quality of our obtained data. 

In \citet{kamath14}, we established the SMC membership of
\objectname[J005252.87-722842.9]{J005252} using low-resolution
spectra. In this study, we used our optical high-resolution spectra and confirmed the
SMC membership of J005252. Using the Fourier cross-correlation technique
we derived its heliocentric radial velocity to be
149\,$\pm$\,1\,kms$^{-1}$. We note that for the cross-correlation we
used the neutral oxygen lines, the O\,I triplet, at 6155.97, 6156.78 and 6158.19\,\AA. The estimated heliocentric radial velocity agrees well with the
velocity ($\approx$\,160\,kms$^{-1}$) expected for stars in the SMC \citep[][]{depropris10}.




\section{Spectral analysis}
\label{spec_anal}

To study the photospheric composition of \objectname[J005252.87-722842.9]{J005252}, we performed a
systematic spectral analysis. We first estimated accurate atmospheric parameters for this object
and then we carried out a detailed abundance study. The
spectral analysis was performed using PyMoog, our own Python wrapper
around the local thermal equilibrium (LTE) abundance calculation
routine MOOG \citep{sneden73}. For the spectral analysis we used the LTE Kurucz-Castelli
atmosphere models \citep{castelli03} and the spectral lines were 
identified using linelists, ranging from 3000 to 11000\,\AA, from the VALD database \citep{kupka99}
combined with the linelist of 'Instituut voor Sterrenkunde' 
\citep{vanwinckel00}. Full details of PyMoog and our spectral
analysis are presented in \citet[][and references therein]{desmedt15}. In the following subsections we briefly describe
our spectral analysis and the obtained results. 

\subsection{Atmospheric parameter determination}
\label{atmos_param}

The atmospheric parameters of \objectname[J005252.87-722842.9]{J005252} were determined 
using both \feI\, and \feII\, lines combined with standard spectroscopic
methods. We determined the effective temperature (\teff) by requiring the
iron abundance to be independent of lower excitation potential. Since 
the low-resolution spectroscopic analysis of \objectname[J005252.87-722842.9]{J005252} carried out in
\citet{kamath14} as well as the high-resolution high S/N spectra
(available as online supporting information) revealed the object to be of spectral-type A, 
we chose the \feII\, lines (since at these temperatures most of the Fe
is present as \feII\, than \feI) for the temperature determination. 
We determined the surface gravity (\logg) by 
imposing ionisation equilibrium between the individual \feI\, and \feII\,
abundances. The microturbulent velocity (\mv) was determined by requiring the iron
abundance to be independent of the reduced equivalent width
(EW/$\lambda$). Similar to the effective temperature determination, we chose
\feII\, lines for the microturbulent velocity determination. 

We decreased the parameter steps of the Kurucz atmosphere
models using linear interpolation to calculate atmospheric models
which lie within the parameter steps. We did not do any extrapolation. 
We chose \teff\, steps of 125\,K, \logg\, steps of
0.25\,dex and \mv\, steps of 0.25\,kms$^{-1}$. 
The
atmospheric parameter and radial velocity estimates for \objectname[J005252.87-722842.9]{J005252} are listed in
Table~\ref{table:atmos_param}. The estimated abundances of \feIh\, and
\feIIh\, differ by 0.10\,dex (see Table~\ref{table:atmos_param}). For warm, low-gravity stars such as
\objectname[J005252.87-722842.9]{J005252}, departures from LTE can be expected to be significant
for many, if not most, elements and transitions. Such non-LTE (NLTE) effects on
the level populations are largely driven by the intense radiation field at shorter
wavelengths and are not compensated by thermalising collisions in the low density
atmosphere of \objectname [J005252.87-722842.9]{J005252}. 
Neutral, relatively low-ionisation minority species such as \feI\, 
are affected by over-ionisation which makes the lines weaker while the overall 
population of FeII, being the dominant ionisation stage, is well approximated 
by the Saha distribution \citep[e.g.,][]{asplund05a}. This difference makes it important
to take NLTE effects into account for the surface gravity determination in particular.
For the purpose we have made use of the extensive 1D non-LTE calculations for Fe by
\citet{amarsi16}; the employed
Fe model atom is described further in Lind et al (in preparation) but we note that
we make use of new, realistic quantum mechanical-based cross-sections for
inelastic collisions between Fe and H (Barklem et al., in preparation). For
\objectname[J005252.87-722842.9]{J005252} we find typical non-LTE effects for \feI\, of 0.18\,dex while \feII\, is essentially unaffected
($\le 0.01$\,dex). We note that while absolute abundances can be sensitive to the
adopted model atmospheres, the NLTE abundance corrections are much less so,
and therefore consider the inconsistency between the Kurucz-Castelli model atmospheres 
used for the LTE abundance analysis and the MARCS (Gustafsson et al. 2008) models
for the NLTE calculations to be of secondary importance compared with other
uncertainties in the observations and analysis. We
assume the estimated \feIIh\, abundance of the object as its metallicity (\feh). We find that this object
is metal poor (\feh\,$=$\,$-$1.18\,$\pm$\,0.10) when compared to
the mean metallicity 
(\feh\,$\approx$\,$-$0.7) of the young
stars in the SMC \citep{luck98}. The indicated uncertainties of
\feIh\, and \feIIh\, are the total errors
on the iron abundances which includes line-to-line scatter and the
atmospheric parameter uncertainties \citep[see][for full details on
the error estimation]{desmedt15}. In Table~\ref{table:atmos_param} we present a
comparison of the estimated atmospheric parameters of
\objectname[J005252.87-722842.9]{J005252} from this study to those 
obtained using the low-resolution AAOmega optical spectra 
\citep{kamath14}. The values from the high-resolution (UVES)
  and low-resolution (AAOmega) spectral analysis are mostly in good
  agreement however discrepancies in, for e.g., the [Fe/H] values, indicate the need for
high-resolution spectra to get reliable stellar parameters and
abundances.

\subsection{Reddening, luminosity and initial mass estimates}
\label{lum+ini_mass}

A main characteristic feature of a post-AGB star is its mid-IR
excess. This large mid-IR excess results in a significant reddening,
which included both the foreground extinction (see Section~\ref{phot}) as
well as the circumstellar reddening. Therefore, it is important that
we estimate the total reddening for
\objectname[J005252.87-722842.9]{J005252}. To do this we follow a
similar method used in \citet{kamath14}. We calculated
the E($B-V)$ by estimating the value of E($B-V)$ that minimised the sum of the 
squared differences between the de-reddened observed and the intrinsic 
$B$, $V$, $I$ and $J$ magnitudes. We used the extinction law by
\citet{cardelli89}, with a Rv\,=\,3.1. We note that we included the uncertainty of the
model atmosphere parameters (such as \teff\, and \logg\,) in our minimisation. At longer wavelengths, emission from 
dust can contribute to the observed magnitudes. We note that, it is possible
that the circumstellar extinction law is different from the interstellar extinction law but we have not explored this
possibility. We derived a small  E($B-V$) value of 0.02\,$\pm$\,0.02.
The derived E($B-V$) values were used to correct the observed magnitudes for extinction. 
Then the $BVIJ$ fluxes of the best-fit (derived from the spectral
analysis) LTE Kurucz-Castelli model atmosphere \citep{castelli03} were normalised 
to the corrected $BVIJ$ fluxes. We note that the uncertainty on the reddening was computed by determining the confidence intervals of the free parameters.

For post-AGB stars, the central 
star is surrounded by circumstellar dust that is not necessarily spherically symmetric. For such cases, the 
observed luminosity $L_{\rm obs}$ (obtained by integrating the flux under
the observed SED, see Section~\ref{phot}) could either be over-estimated or under-estimated. For this reason 
it is essential to estimate the photospheric luminosity of the central
star $L_{\rm ph}$.  We determined this luminosity by integrating the model atmosphere scaled to the dereddened data. We estimated a
photospheric luminosity of 8200\,$\pm$\,700\,\Lsun. The
uncertainty on the luminosity is dominated by the uncertainty of the
total line-of-sight reddening, 
which impacts strongly on the luminosity determination. 

In Table~\ref{table:atmos_param} we present the SED results of
\objectname[J005252.87-722842.9]{J005252}. The newly constructed SED
is shown in Figure~\ref{J005252_sed}. 

The determined luminosity and reddening agrees well with the values
obtained using the low-resolution spectra for this object
\citep{kamath14}.

Using the derived \teff, \feh, and luminosity ($L_{\rm ph}$), we
determined the current mass of \objectname[J005252.87-722842.9]{J005252} to
be 0.63\,$\pm$\,0.02\,\Msun\, using the luminosity-core mass relation for AGB stars \citep[][the
relation between core mass and quiescent luminosity, equation\,(3)]{wood81}.  We also determined the initial mass to be
~1.5\,\Msun\, to 2.0\,\Msun\, using the initial-final-mass relation
(corresponding to a Z\,=\,0.001) of 
\citet{vw94}. Models by \citet{fishlock14} yields a mass of 
$\sim$\,0.64\,$-$\,0.65\,\Msun from the luminosity of a model
corresponding to a star near the tip of the AGB with Z\,$=$\,0.001 and
an initial mass of 
$\sim$\,2\,\Msun. The Z\,=\,0.001 FRUITY models by
\citet{cristallo15}, the COLIBRI models by \citet{marigo13} and the models by \citet{ventura14} yield a luminosity-core
mass relationship that is consistent with \citet{fishlock14} at
~2.0\,\Msun. 

We note that there are uncertainties in these
  estimates which arise from depend the models used and the
assumptions made in the modeling. To complicate this 
further there is considerable degeneracy in the model predictions around
this mass range. The core-mass luminosity relation may give us a core-mass
but working backwards to get the initial mass suffers significantly 
from this degeneracy.

\subsection{Establishing the post-AGB nature of J005252}
\label{pagb}

Our surveys for post-AGB stars in the MCs \citep{kamath14,kamath15}
revealed that other classes of objects, such as core-He burning (CHeB) stars
and pre-main sequence (PMS) stars  can appear similar to post-AGB
stars based on either their luminosities or dust excesses. 

\begin{figure}
\epsscale{.80}
\plotone{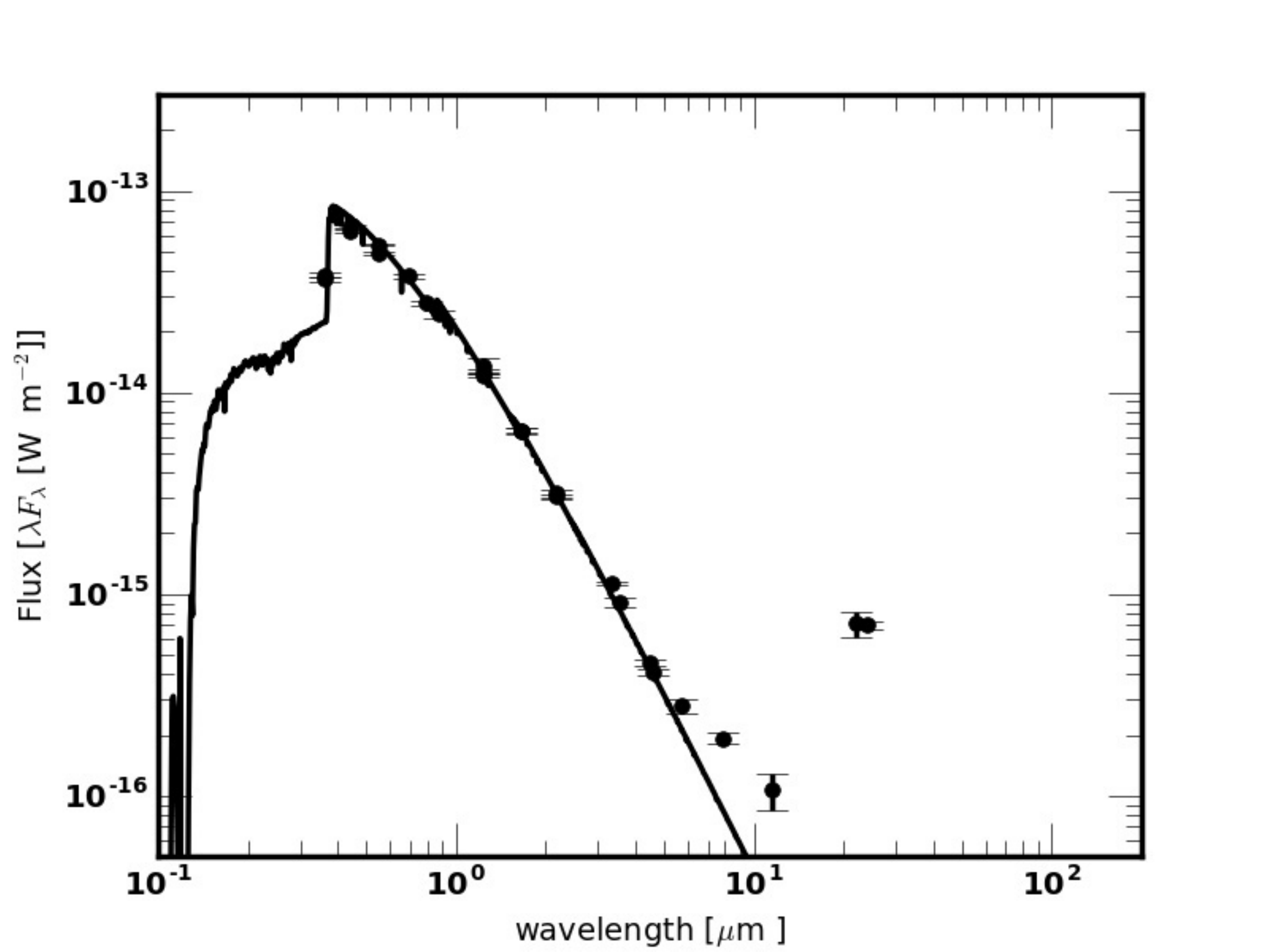}
\caption{Spectral energy distribution for J005252. The black symbols represent the
  dereddened photometry. The black line represents the best-fitting
  scaled Kurucz model atmosphere (see text for details). \label{J005252_dered_sed}}
\end{figure}

CHeB stars (of 6\,$-$\,10\,\Msun) have
similar luminosities as
\objectname[J005252.87-722842.9]{J005252}. However, the CHeB phase is
not predicted to involve heavy mass loss
and hence stars in this phase should not have significant
circumstellar dust emission. This is contrary to the situation for 
\objectname[J005252.87-722842.9]{J005252} where the object shows an
excess already at 8 microns (of $\sim$\,0.8 magnitudes) which
continues up to 24\,$\mu$m (see
Figure~\ref{J005252_dered_sed}). There is no information available on the
mid-IR excess beyond 24\,$\mu$m. In the large sample of LMC Cepheids studied by \citet{neilson09}, 
only $\sim$\,2\,$\%$ have 8 micron excesses of this size or
greater. Also, our object has a
  relatively low \logg\, value. If this object was a CHeB star then it would be $\sim$\,8\,$-$\,12 times more massive than a
  post-AGB star at the same luminosity \citep[based on
  evolutionary tracks of][for Z\,=\,0.001]{bertelli08}. They would
  therefore have a \logg\, larger by $\sim$\,1\,dex. However this is
  not the case.  Additionally, the metallicity of
  \objectname[J005252.87-722842.9]{J005252}
  (\feh\,=\,$-$1.18\,$\pm$\,0.10) is lower than that expected from
  CHeB stars with \feh\,$=$\,$-$0.7\,$\pm$0.15 \citep[based on a study
  of 12 cepheids by][]{romaniello05}. Based on these arguments, 
\objectname[J005252.87-722842.9]{J005252} is not likely to be a CHeB
  star. 

Though PMS stars can have mid-IR excesses and luminosities
similar to \objectname[J005252.87-722842.9]{J005252}, at a given
luminosity, the mass of such a PMS star is about $\sim$\,15\,$-$\,20 times that of
the corresponding post-AGB star. This leads to a difference of $\sim$\,1.2
in \logg\, between a post-AGB star and a PMS star at a given
luminosity \citep[based on PMS evolutionary tracks of][]{tognelli11}. This gravity difference makes it more likely for
\objectname[J005252.87-722842.9]{J005252}  to be a post-AGB star than
a PMS star. Furthermore, the estimated  metallicity (see Section~\ref{atmos_param}) of
  \objectname[J005252.87-722842.9]{J005252} is much lower
  (Z\,$\approx$\,0.001) than the typical metallicity of young stars in
  the SMC (Z\,$\approx$\,0.004). Hence if this object was to be a PMS 
star then it should have had the metallicity of the young
  stars in the SMC. This implies that our object is of lower mass
  (older) than PMS stars of the same luminosity. 

Therefore, we conclude on balance that \objectname[J005252.87-722842.9]{J005252}
is likely a post-AGB star. 

\subsection{Abundance Analyses}
\label{abund_anal}

We used the atmospheric parameters listed in
Table~\ref{table:atmos_param} to calculate detailed abundances for
\objectname[J005252.87-722842.9]{J005252}. We note that we analysed, in detail, the entire available
spectral region (from 3300 to 4220\,\AA, 4790 to
5750\,\AA\, and 5840 to 6800\,\AA) but used only isolated non-blended lines for our abundance
analysis.  We did not use lines with EWs smaller than 5\,m\AA\, because as they may
be confused with spectral noise, nor larger than 150\,m\AA, because these
are saturated and formed in the upper layers of the photosphere where
non-LTE effects are likely larger.

Based upon the ionisation potential of the corresponding ion, the
element over iron ratios ([X/Fe]) were calculated using \feI\, or \feII. If the ionisation potential of an
ion is below the ionisation potential of \feI, then the abundance of
\feI\, is used for calculating [X/Fe]. If the ionisation potential exceeds
the ionisation potential of \feI, \feII\, was used for calculating
[X/Fe]. The same principle was also used for calculating the total
error of [X/Fe]. The errors were determined using the method described
in \citet{deroo05a}. The uncertainties in the abundances, due to the atmospheric 
parameters, were calculated by determining the abundances of a certain
ion for atmospheric models with \teff\,$\pm$\,125\,K, models with
\logg\,$\pm$\,0.25\,dex, and with microturbulent velocity 
\mv\,$\pm$\,0.2\,kms$^{\rm -1}$.  The total uncertainties are therefore the
quadratic sum of the uncertainties of the mean due to line-to-line
scatter ($\sigma_{\rm l2l}$), uncertainties due to atmospheric
parameters and the Fe abundance
uncertainty ($\sigma_{\rm Fe}$), \citep[see][for further details on
the error analysis]{desmedt15}. 

\begin{figure}
\begin{center}
\epsscale{1.1}
\plottwo{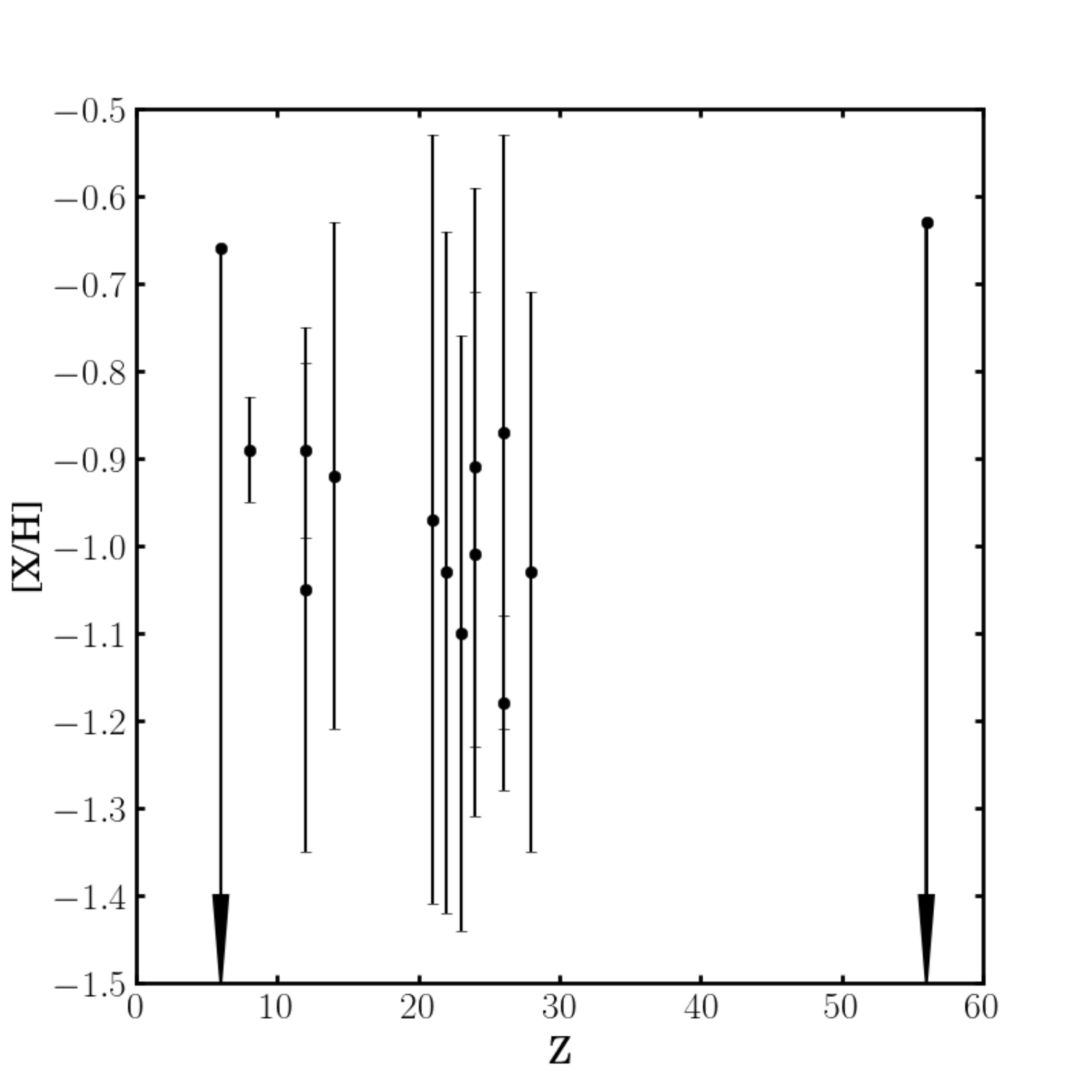}{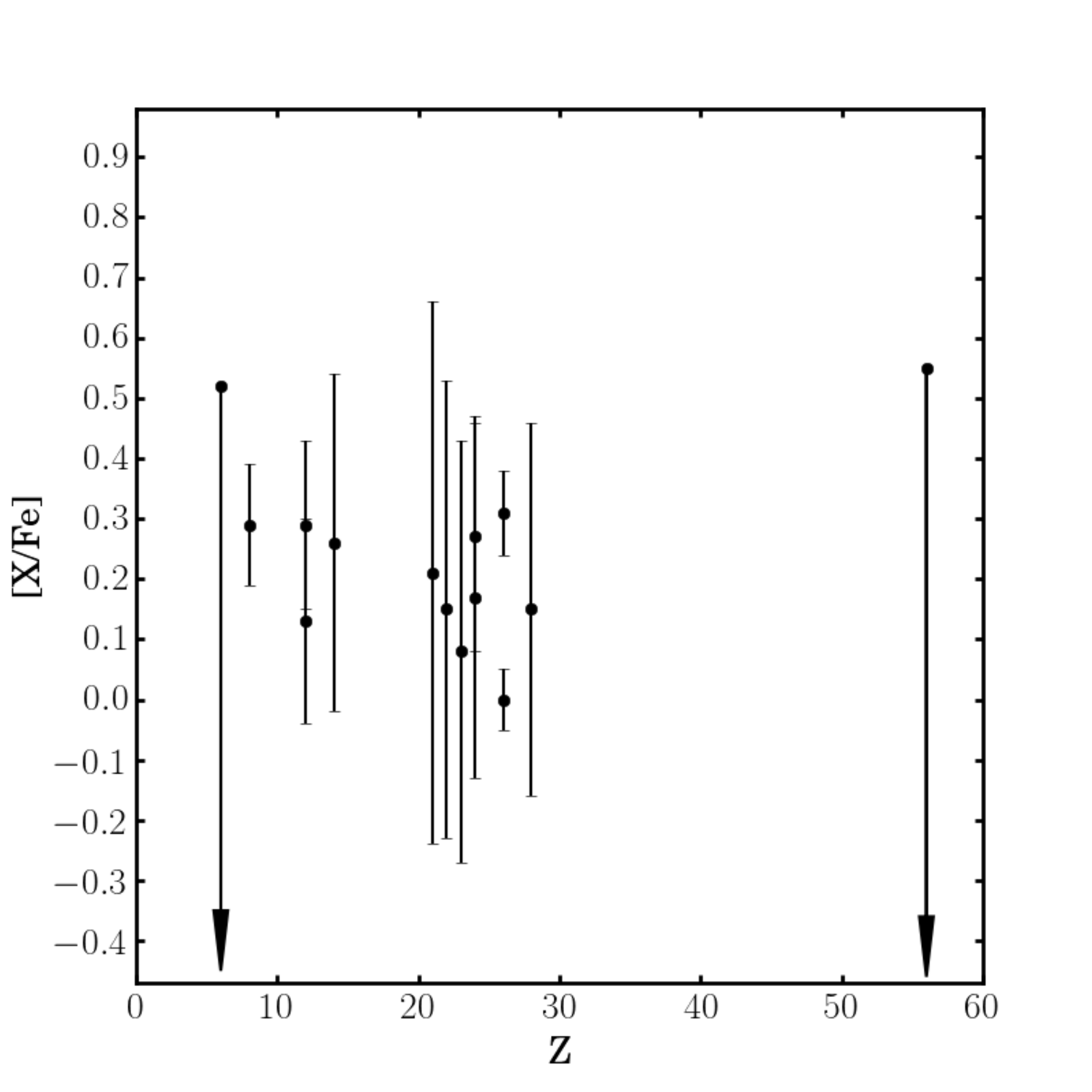}
\caption{Left panel: element over hydrogen [X/H] ratios of J005252. The error bars respresent the
  total uncertainties $\sigma_{\rm tot}$ in [X/H]. Right panel:
  element over iron [X/Fe] ratios of J005252. The errors bars respresent the
  total uncertainties $\sigma_{\rm tot}$ in [X/Fe]. Note: the abundances of C\,(Z\,=\,6) and Ba\,(Z\,=\,56) are
upper limits and marked with downward
arrows.\label{J005252_abund}. For some of the elements we were able to
estimate abundances using both neutral and singly-ionised
lines. For such cases, we show both the abundances. For O\,I and Fe\,I
we plot their abundances corrected for NLTE effects (see Table~\ref{table:abund_anal}).}
\end{center}
\end{figure}

The spectra are devoid of lines and the only elements that were
identifiable from the spectra of \objectname[J005252.87-722842.9]{J005252} are
O\,(Z\,=\,8), Mg\,(Z\,=\,12), Si\,(Z\,=\,14), Sc\,(Z\,=\,21), Ti\,(Z\,=\,22), V\,(Z\,=\,23), Cr\,(Z\,=\,24), 
Fe\,(Z\,=\,26), and Ni\,(Z\,=\,28) and for these
elements we determined their elemental abundances. 
The results of our abundance analysis are summarised in
Table~\ref{table:abund_anal} and are also graphically presented in
Figure~\ref{J005252_abund}.  We take into account NLTE effects for \feI\, (see
Section~\ref{atmos_param}) and for O\,I (see below). Unfortunately few NLTE 
studies of other elements for stellar parameters similar to \objectname[J005252.87-722842.9]{J005252} have been performed to
date and we therefore have to settle for the 
LTE abundances; we stress that there can be substantial systematic errors for
individual elements as a result of this simplifying assumption. We encourage dedicated
non-LTE calculations for key elements. 

In table~\ref{table:linelist} we list all the spectral lines used
in our abundance analysis. We note that the full table is available as online supporting
information.

Our analyses reveals that \objectname[J005252.87-722842.9]{J005252}
has a rather intriguing abundance distribution. 
It is relatively metal poor (with \feh\,=$-$1.18\,$\pm$\,0.10) with
[O\,I/Fe]\,=\,0.75\,$\pm$\,0.1 and [O\,I/H]\,=\,$-$\,0.89\,$\pm$\,0.06. We take into account non-LTE effects for
O\,I using the 1D non-LTE computations by \citet{amarsi16}. For the 
O\,I 6158\AA\, triplet, the non-LTE effects are very substantial in 
\objectname[J005252.87-722842.9]{J005252}: $-0.41$\,dex. 

\begin{figure}
\begin{center}
\epsscale{1.10}
\plottwo{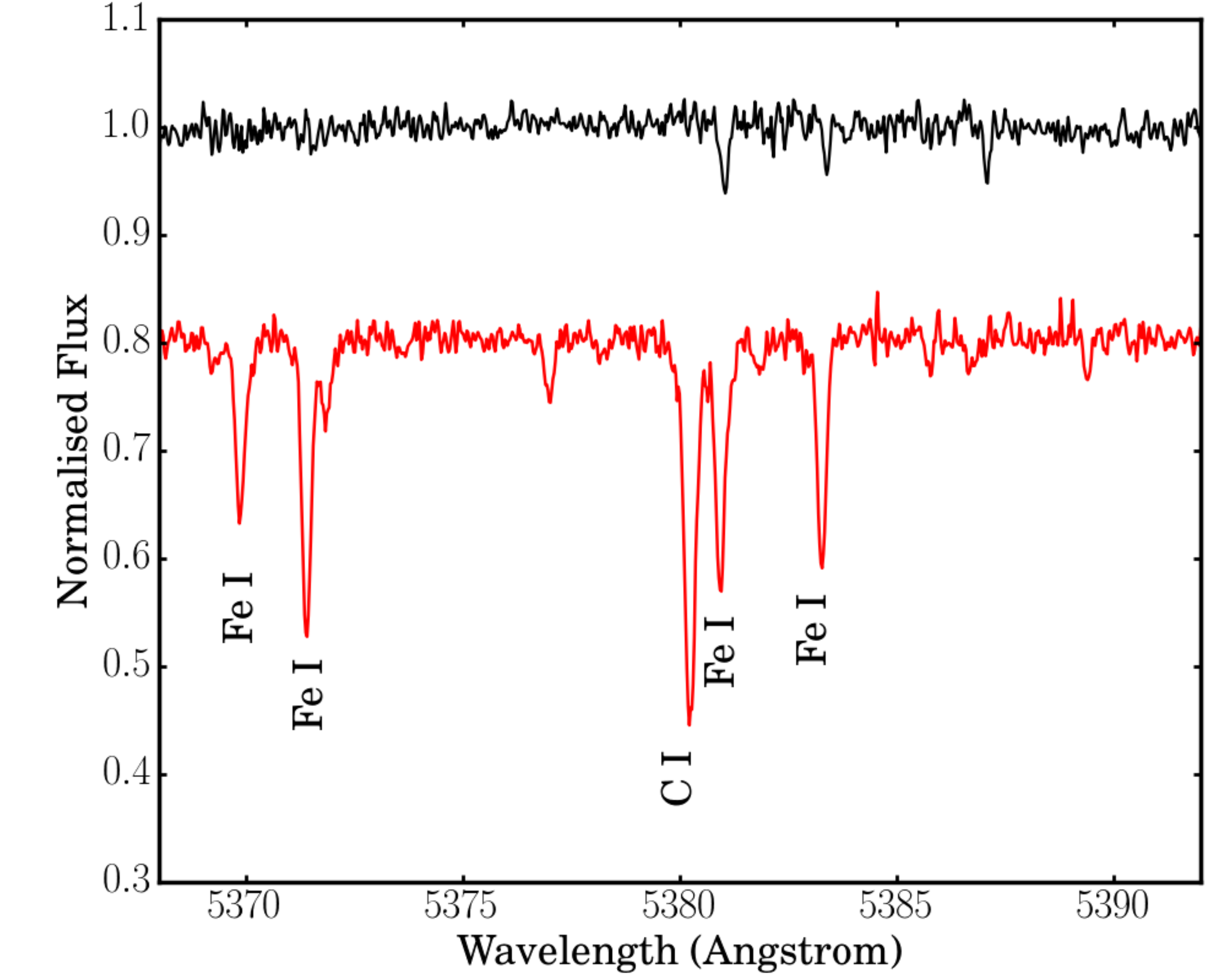}{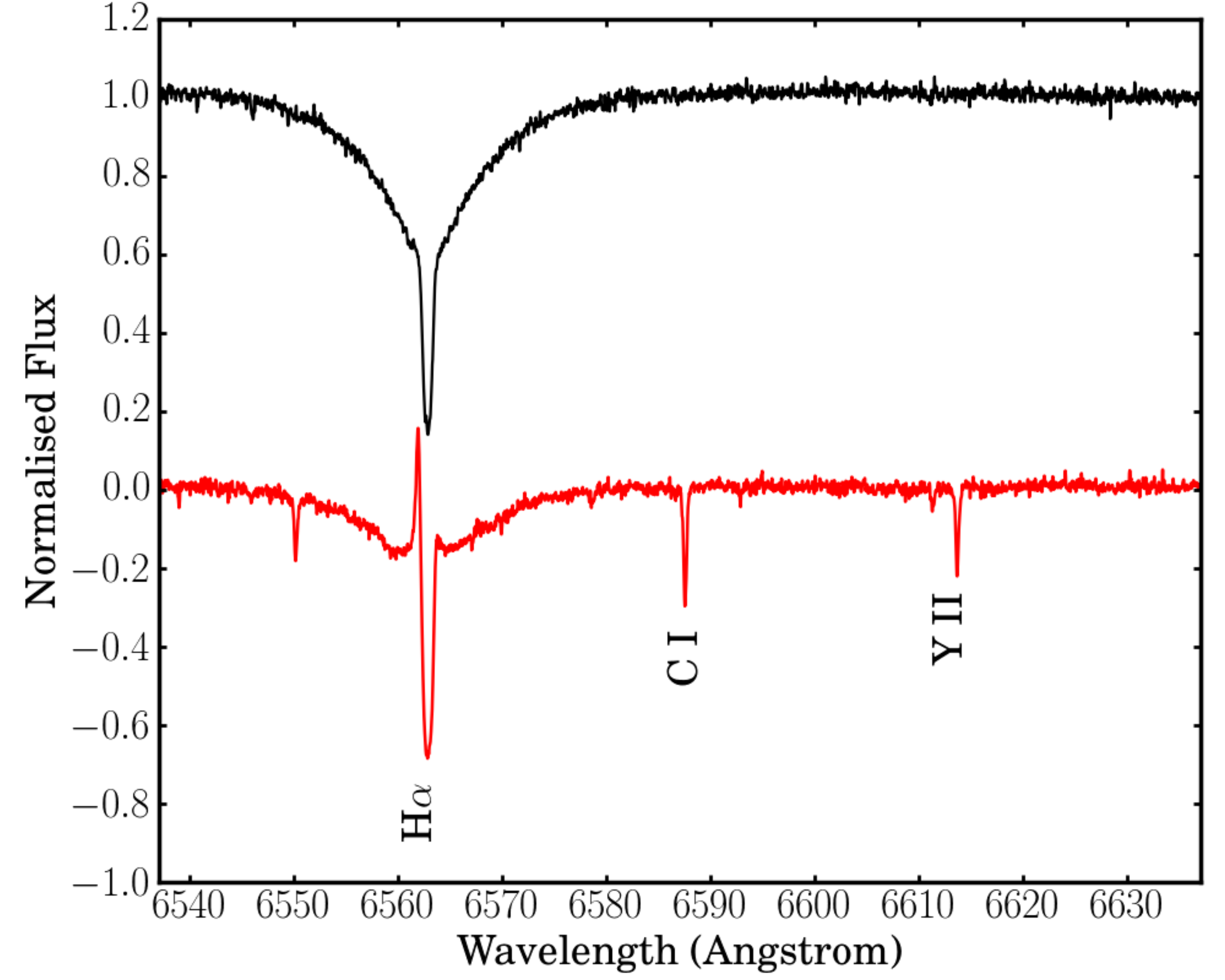}
\caption{Normalised spectra of J005252 (in black) covering the region that
  contains strong neutral carbon (C\,I) lines at 5380.34\,\AA\, (left panel) and
  6587.62\AA, (right panel). The spectra of J050632.10-714229.8 (in
  red), a known $s$-process enriched, single LMC post-AGB star, with
  similar stellar parameters to J005252, is shown for comparison. See
  text for further details. \label{J005252_nocarbon}}
\end{center}
\end{figure}

\objectname[J005252.87-722842.9]{J005252}
shows no signs of C-enhancement. In our spectral coverage, the
strongest usable lines of C are the neutral carbon (C\,I) lines at 5380.34\,\AA\, and
6587.62\,\AA. However, neither of the two
lines were detected above the noise. Figure~\ref{J005252_nocarbon}
shows the spectral regions that cover the 5380.34\,\AA\, and the
6587.62\,\AA\, C\,I lines. We used both lines to derive an upper limit for C which yielded a [C/Fe] of 0.3, for an assumed EW of 5\,m\AA. We were unable
estimate the abundance of N or provide an upper limit for
the abundance as the strongest lines of N lie outside our spectral
coverage. 

\begin{figure}
\begin{center}
\epsscale{1.10}
\plottwo{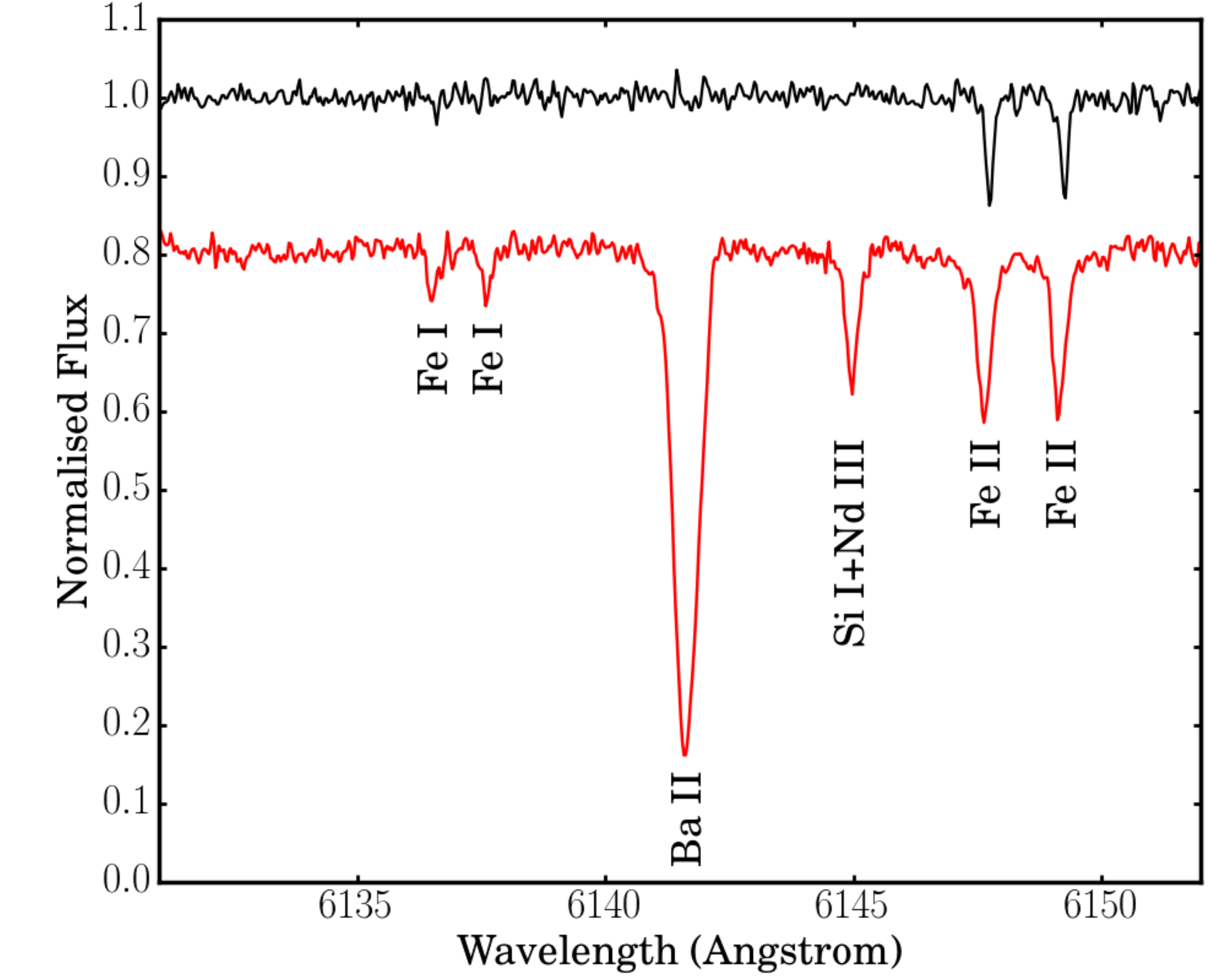}{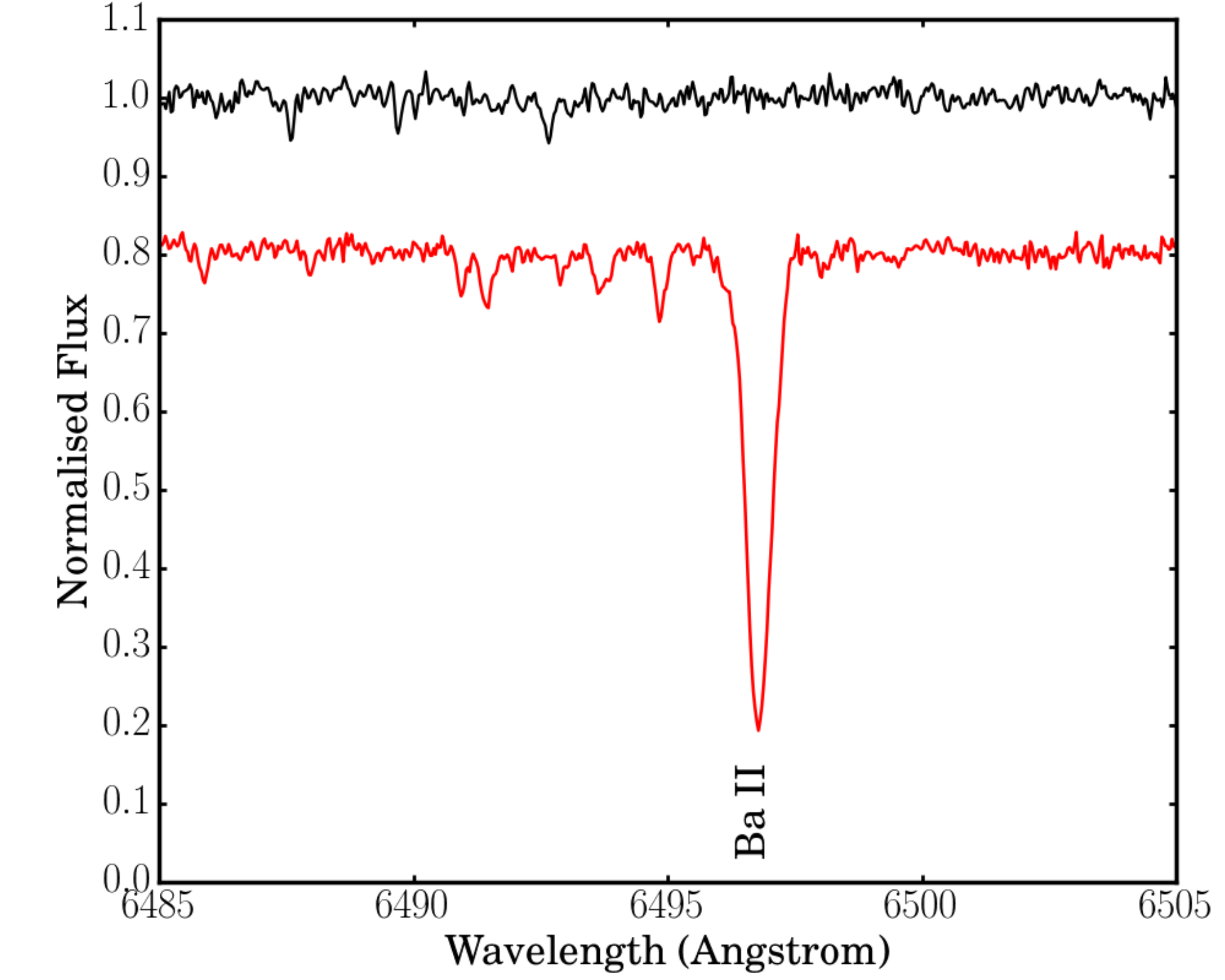}
\caption{Normalised spectra of J005252 (in black) covering the region that
  contains strong singly ionised barium (Ba\,II) lines at
  6141.72\,\AA\, (left panel) and
  6496.90\AA\, (right panel). The spectra of J050632.10-714229.8 (in
  red), a known $s$-process enriched, single LMC post-AGB star, with
  similar stellar parameters to J005252, is shown for comparison. See
  text for further details.\label{J005252_nobarium}}
\end{center}
\end{figure}

\objectname[J005252.87-722842.9]{J005252} also showed no signs of $s$-process
enrichments. Stars that are $s$-process enriched show the presence of
light $s$-elements (ls-elements) such as yttrium (Y, Z=39) and
zirconium (Zr, Z=40), and heavy $s$-elements (hs-elements), such as
barium (Ba, Z=56), lanthanum (La, Z=57), cerium (Ce,Z=58), praesodymium (Pr, Z=59),
neodymium (Nd, Z=60) and samarium (Sm, Z=62) \citep[see][and
references
therein]{desmedt12,vanaarle13,desmedt15}. Figure~\ref{J005252_nobarium}
shows the spectral regions that cover the strongest Ba\,II line at 6141.72\,\AA\, and a weaker line at 6496.90\,\AA. It can
be clearly seen that these Ba lines are absent in the spectra
therefore implying that 
\objectname[J005252.87-722842.9]{J005252} is
not $s$-process enriched. We used the strongest Ba\,II line at
6141.72\,\AA\, to determine an upper limit for
Ba. We obtained a [Ba/Fe]\,$<$\,0.55, for an assumed EW of 5\,m\AA, which is above
solar. We note that there is a mild $s$-process enrichment in
the initial compostion of the LMC, starting from solar value at
\feh\,=\,$-$1.5 and reaching 0.8 for
\feh\,$\geq$\,$-$0.3 \citep{vanderswaelmen13}. With the 
assumption that the SMC may also be mildly $s$-process enriched, it is
very likely that
\objectname[J005252.87-722842.9]{J005252} is not $s$-process
enriched. 

\begin{figure}
\begin{center}
\epsscale{1.10}
\plottwo{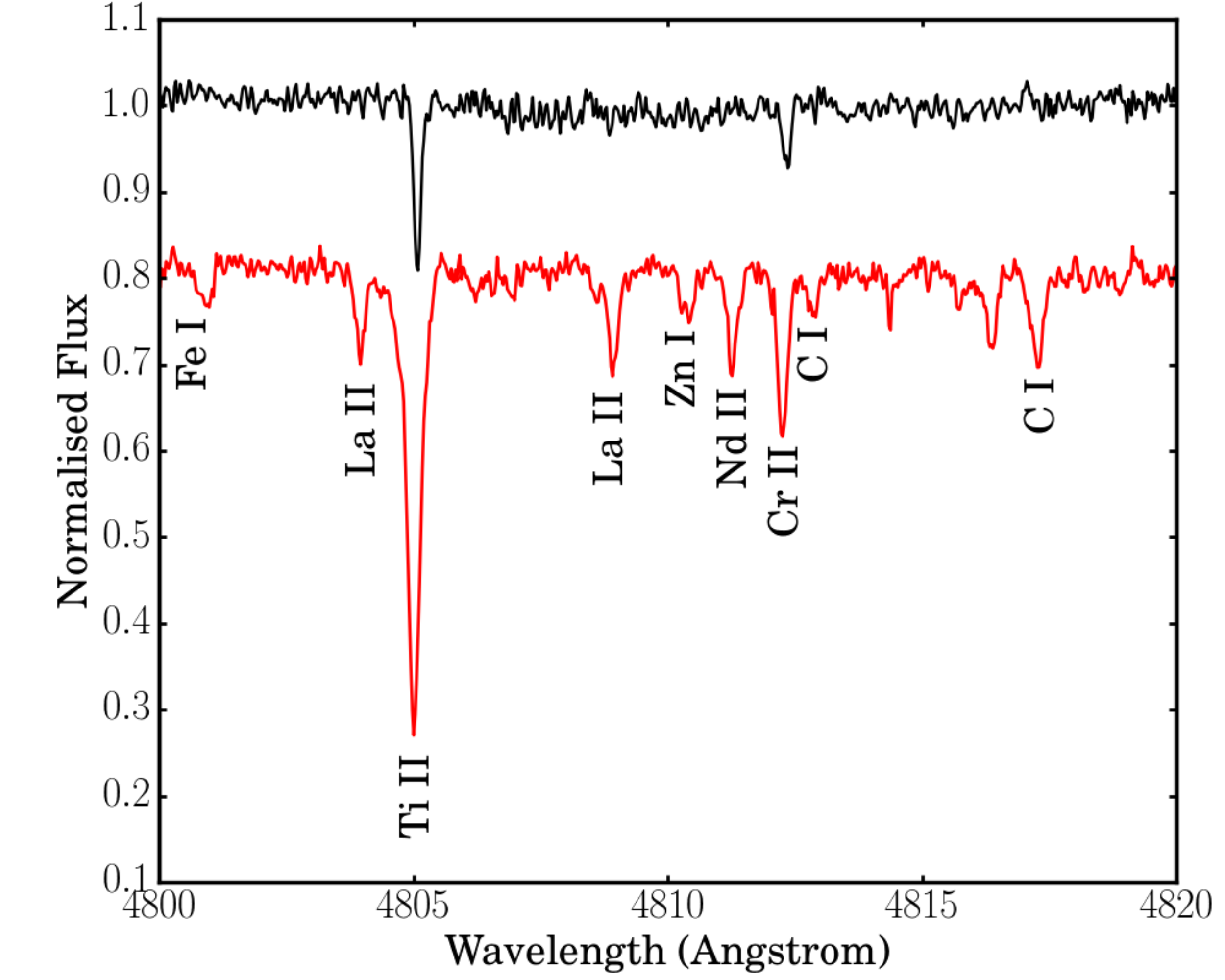}{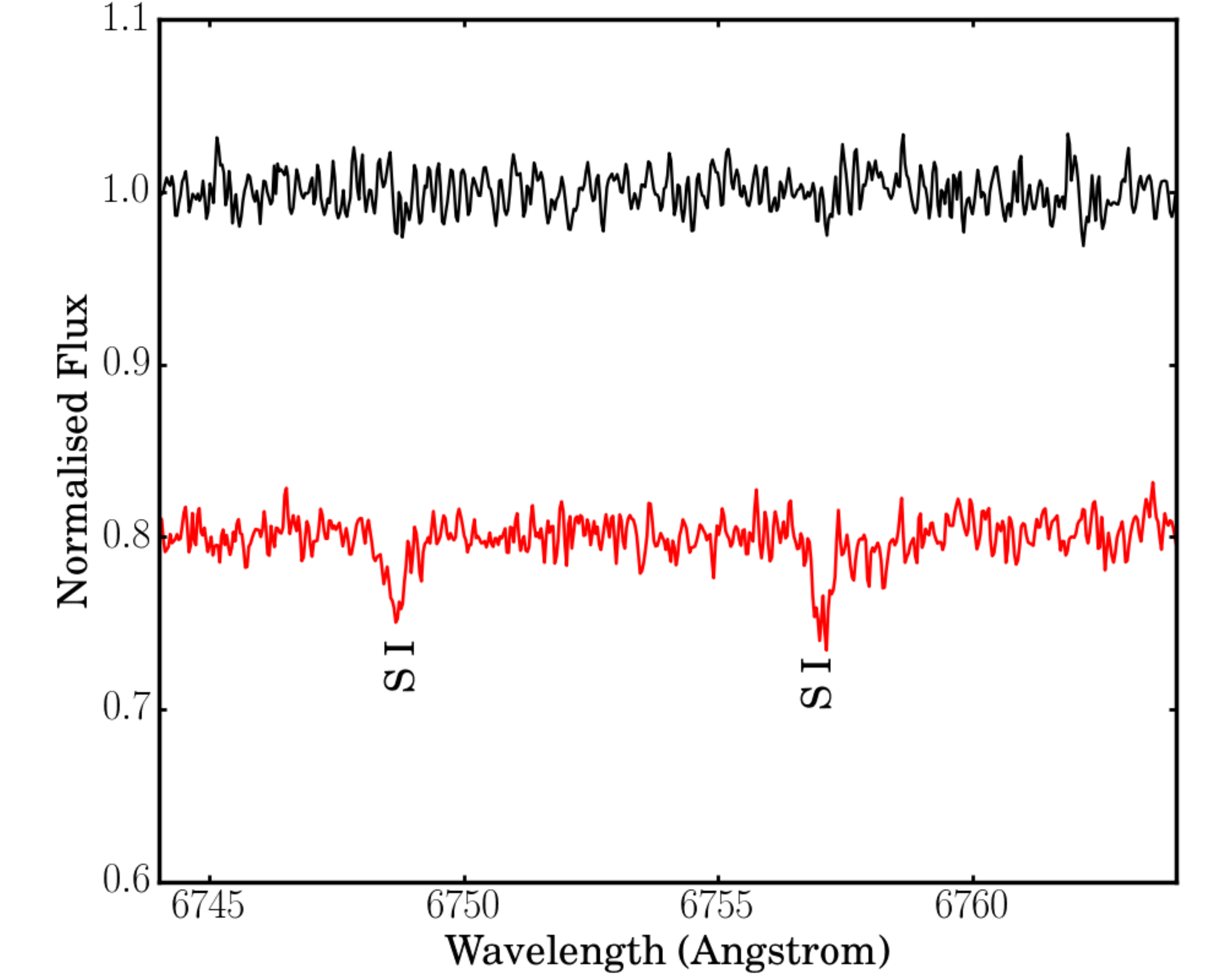}
\caption{Normalised spectra of J005252 (in black) covering the region that
  contain a strong neutral zinc (Zn\,I) line at 4810.50\,\AA (left
  panel) and neutral sulphur (S\,I) lines at 6757.16\,\AA and 6748.79\,\AA (right panel). The spectra of J050632.10-714229.8 (in
  red), a known $s$-process enriched, single LMC post-AGB star, with
  similar stellar parameters to J005252, is shown for comparison. See
  text for further details.\label{J005252_no_Zn_S}}
\end{center}
\end{figure}

Another
chemical trend displayed by post-AGB stars is 'depletion'. Affected
photospheres are characterised by a selective depletion of refractory
elements, while the volatile elements retain their original abundance
(see Section~\ref{intro}). The best elements to trace depletion are Zn
and S \citep{vanwinckel03}. In Figure~\ref{J005252_no_Zn_S}, we show the spectral
regions that cover the Zn\,I line at 4810.50\,\AA\, and the S\,I lines
at 6757.16\,\AA\, and 6748.79\,\AA. It can be seen that these lines
cannot be detected in the spectra of 
\objectname[J005252.87-722842.9]{J005252}. At the \teff\, of 
\objectname[J005252.87-722842.9]{J005252} we do not expect to see
Zn\,I. Therefore the absence of the
strongest Zn\,I line at 4810.50\,\AA\, does not yield a useful upper
limit for Zn. We used the strongest useful S\,I lines at 6757.16\,\AA\, 
and 6748.79\,\AA\, to quantify an upper limit for the S abundance. We
estimated an upper limit of [S/Fe]\,$\leq$\,1.0. The
estimated upper limit for the S abundance and the lack of detectable
S\,I lines (as show in Figure~\ref{J005252_no_Zn_S}) implies that 
\objectname[J005252.87-722842.9]{J005252} does not show signs of
depletion. 

\section{Understanding the photospheric chemistry of J005252}
\label{discussion}

Detailed chemical abundance studies of post-AGB stars have shown that 
they are chemically much more diverse than anticipated and most of
these chemical diversities have been tied to the evolutionary channels
of these stars. For instance, some of the single post-AGB objects are
the most C-rich and  $s$-process enriched objects 
known to date while others, the likely binary objects, are not
enriched at all and exhibit a photospheric depletion (see
Section~\ref{intro}). The object of this study, \objectname[J005252.87-722842.9]{J005252}, has a rather intriguing chemistry which
does not comply with any of the known chemical trends observed in post-AGB
stars \citep[see][for a review]{vanwinckel03}, therefore making this object a chemically very peculiar star.  In the following subsections we interpret the
observed photospheric composition of \objectname[J005252.87-722842.9]{J005252}.

\subsection{The carbon abundance of J005252}

Based on the mid-IR excess (due to a detached circumstellar shell
surrounding the central star, see Figure~\ref{J005252_sed}), low
surface gravity (\logg\,=1.0\,$\pm$0.25) and high luminosity ($L_{\rm
  ph}$/\Lsun\,=\,8200\,$\pm$\,700) and low metallicity
(\feh\,=\,$-$1.18\,$\pm$\,0.10) of \objectname[J005252.87-722842.9]{J005252},
this object is very likely to be a post-AGB star. The high luminosity of
the object points to the star having evolved off the TP-AGB phase (see Section~\ref{intro}). 
For this reason we expect the stellar photosphere to be enriched in carbon and
probably enriched with $s$-process elements. However, this is not the case. 
\objectname[J005252.87-722842.9]{J005252} is not C-enhanced (see
Section~\ref{abund_anal} and Figure~\ref{J005252_nocarbon}). There is
no detectable carbon-enhancement. The oxygen is 
above solar with [O/Fe]\,=\,0.29\,$\pm$\,0.1, after taking into
account NLTE corrections. The lack of
C-enhancement in this object indicates that the star did not undergo TDU episodes
or the TDU was inefficient. We could not investigate the nitrogen
abundance as the nitrogen lines lie outside the wavelength range of
our spectra. 

In the SMC, the observed carbon star
luminosity function (CSLF), based on the detected AGB carbon stars,  peaks at an M$_{\rm
  bol}$\,$\approx$\,$-$4.4, \citep[derived using the data presented
in][]{groen04}. \objectname[J005252.87-722842.9]{J005252} (with a
luminosity of 8200\,\Lsun\, corresponds to M$_{\rm bol}$ = $-$5.03) falls within
one sigma of the peak of the CSLF suggesting that it should be enhanced in carbon. 

TP-AGB evolutionary models allow for a wide exploration of the
TDU characteristics as a function of stellar mass, metallicity and mass-loss descriptions
\citep{karakas02,herwig00,herwig04a,herwig04b,weiss09,karakas10a,cristallo11,fishlock14}. 
The predicted surface abundances are, however, subject to many uncertainties as they are not only dependent on the 
TDU physics itself, but also on the dilution of the remaining envelope material and hence on the adopted mass loss description.
The latter will also determine the number of thermal pulses predicted for every model.
In stars less than $\sim$\,2.5\,\Msun, the efficiency of TDU (commonly represented by $\lambda$ which is the ratio between the mass dredged-up into the envelope and the core-mass growth during the interpulse period) has been predicted to increase with stellar mass as well as during the evolution
on the AGB \citep{karakas10a,cristallo11,ventura14,marigo13,fishlock14}.  Lower metallicities favour an earlier onset of TDU and a larger efficiency
resulting in the formation of low-mass C-stars.  Stellar evolution
models \citep{karakas10a,fishlock14} show that for an object such as
\objectname[J005252.87-722842.9]{J005252}, (with an initial mass of
$\sim$\,1.5\,$-$\,2\,\Msun, and Z\,=0.001), the efficiency of TDU
ranges between 0.4 to 0.7 with 10 to 14 TDU episodes with 14 to 17
computed thermal pulses. The COLIBRI model by\citet{marigo13}, (with an initial mass of
$\sim$\,1.5\,$-$\,2\,\Msun, and Z\,=0.001), shows that the efficiency
of TDU reaches a maximum of $\sim$\,0.6 with 9 thermal pulses yielding a
C/O of $\sim$\,10.  For an initial mass of
$\sim$\,1.5\,$-$\,2\,\Msun, and Z\,=0.001, the models by
\citep{ventura14} predict that the efficiency of TDU reaches a maximum
of $\sim$\,0.9 with $\sim$\,11\,$-$\,thermal pulses, yielding a C/O of
$\sim$\,7\,$-$\,13. Therefore, the standard models predict that
\objectname[J005252.87-722842.9]{J005252} should be C-rich. This implies that
\objectname[J005252.87-722842.9]{J005252} should be enhanced in carbon.

Several synthetic models of the TP-AGB phase \citep[e.g.,][]{marigo99,stancliffe05a,izzard04b} have been calculated to fit the
observed carbon star
luminosity function in the SMC. The synthetic models by
\cite{stancliffe05a}, for the SMC metallicity (Z=0.004), find that
carbon stars were formed in all their models with initial masses
between 1\,\Msun\, and 3\,\Msun. Their lowest mass models 
become carbon stars with M$_{\rm
  bol}$\,$=$\,$-$4.2, which is only just below the peak of the SMC CSLF. 

A process that does destroy the production of carbon in AGB stars is
HBB (see Section~\ref{intro}). However, at Z\,=0.001 this is only active in stars
with masses greater than $\sim$\,2.5\,$-$3\,\Msun
\citep{boothroyd93,lattanzio92,lattanzio96,karakas10a,ventura12,fishlock14} and therefore is not likely
to affect the nucleosynthesis of
\objectname[J005252.87-722842.9]{J005252}. 

Therefore based on predictions from both detailed single stellar evolutionary models
as well as synthetic TP-AGB evolution calculations, and based on the stellar parameters ($L_{\rm
  ph}$/\Lsun\,=\,8200, \feh\,=\,-1.18\,$\pm$\,0.10,
$M$/\Msun\,=\,1.5\,$-$\,2\,\Msun, see Section~\ref{spec_anal}), \objectname[J005252.87-722842.9]{J005252} should be C-rich, but the
observations do not agree with the predictions of any model found in
the literature.  

\subsection{The $s$-process abundance of J005252}

\begin{figure}
\begin{center}
\epsscale{1.10}
\plottwo{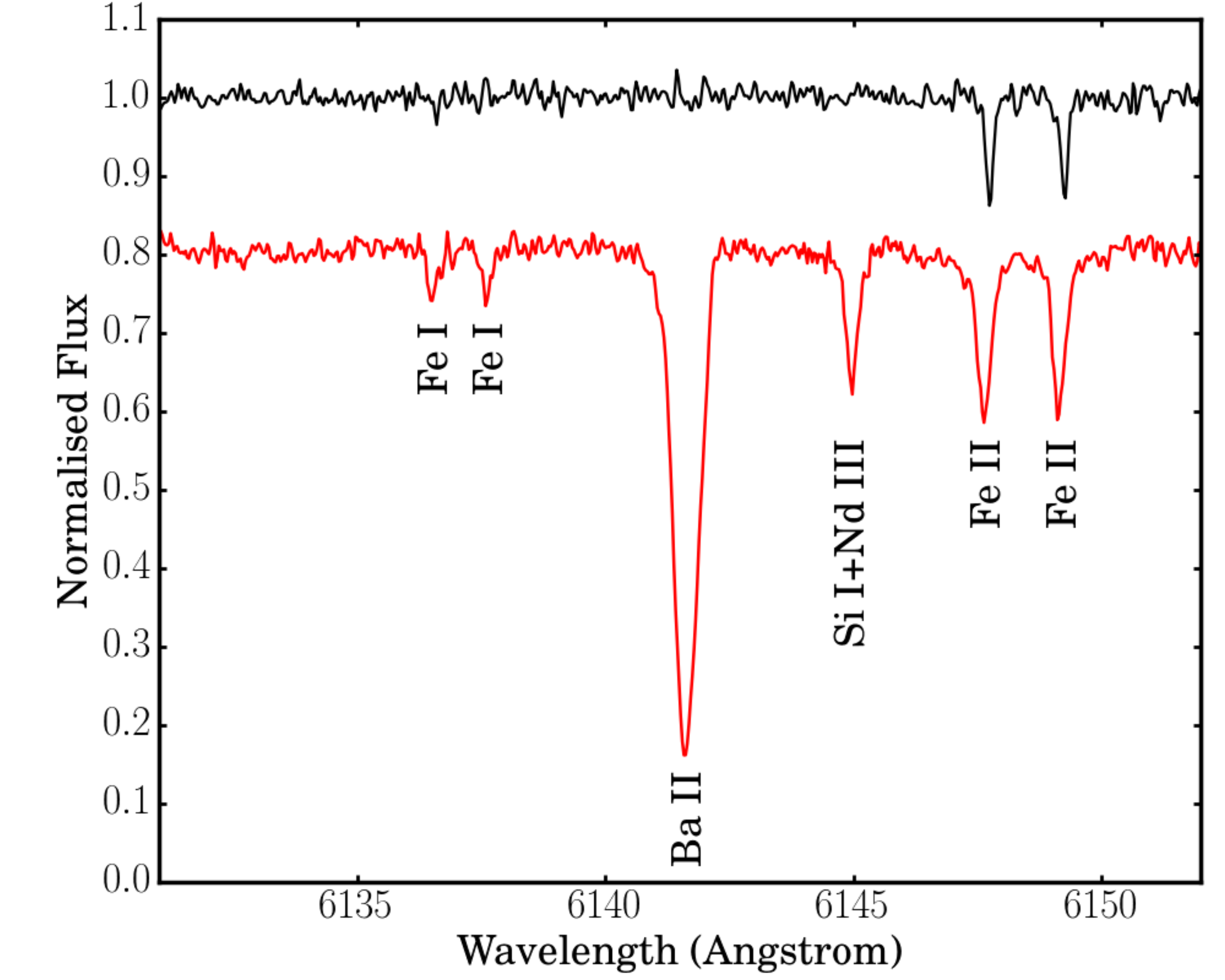}{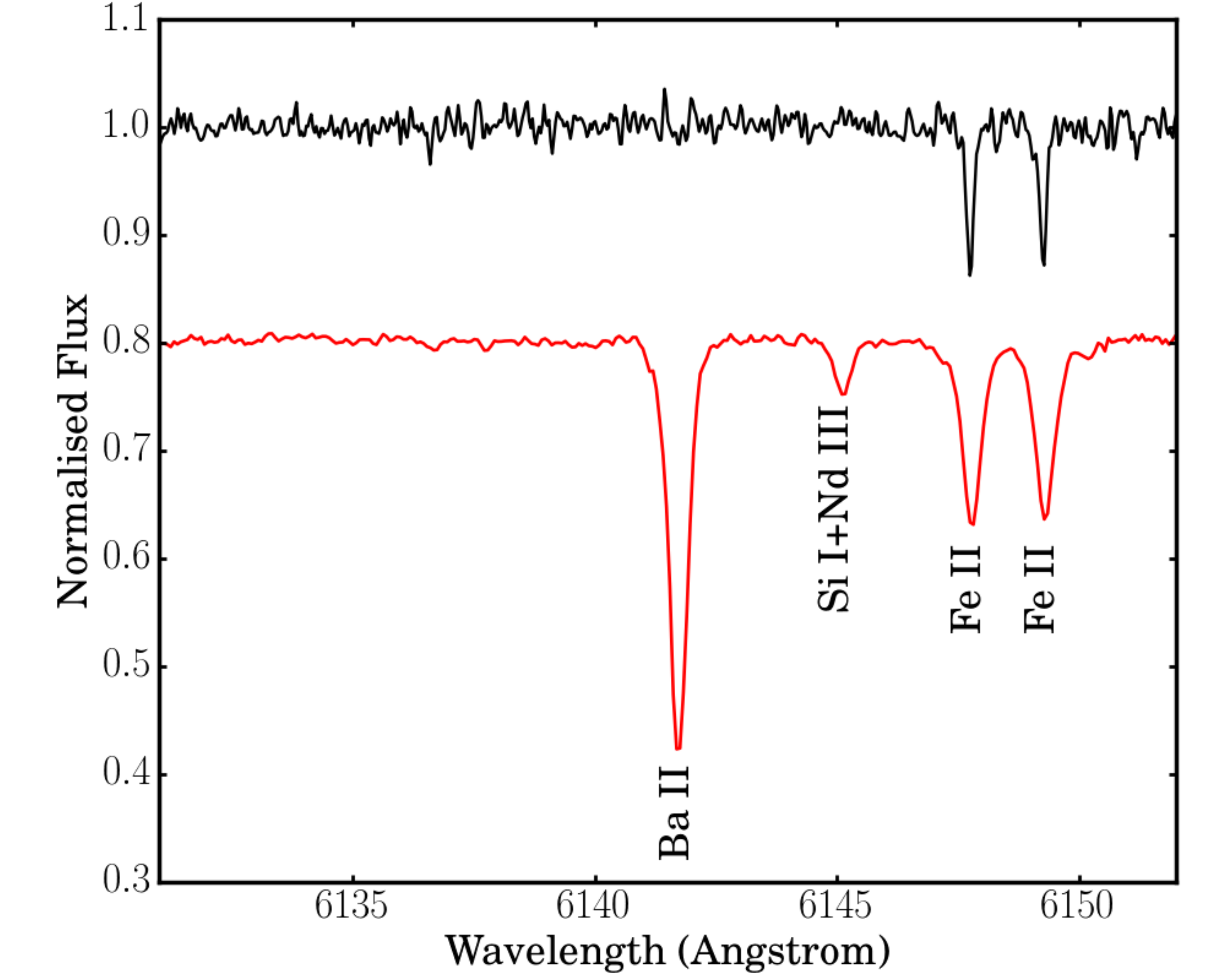}
\caption{ Normalised spectra of J005252 (in black) covering the region that
  contains the singly ionised barium (Ba\,II) line at
  6141.72\,\AA. The left panel shows the spectra of  J050632.10-714229.8 (in
  red), a known $s$-process enriched, single post-AGB star in the LMC with
  similar stellar parameters to J005252. The right panel shows the
  spectra of HD187885 (in
  red), a known Galactic $s$-process enriched, single post-AGB star with
  similar stellar parameters to J00525. See text for more details. \label{comp_s}}
\end{center}
\end{figure}

The SED of \objectname[J005252.87-722842.9]{J005252} is
of shell-type suggesting that this object is a single star. As
mentioned in Section~\ref{intro}, the MC and Galactic
post-AGB stars with shell-type SEDs often show C-enhancement and $s$-process
enrichment. In Figure~\ref{comp_s} we compare the
spectral region covering the Ba\,II line at 6141.72\,\AA\, with the 
$s$-process enriched LMC post-AGB star
\objectname[J050632.10-714229.8]{J050632} \citep{vanaarle13} and the 
$s$-process enriched Galactic post-AGB star \objectname[HD187885]{HD187885}
\citep{vanwinckel96a}. We chose these objects for our spectral comparison as
they have stellar
parameters (\teff, \logg, \feh, see Table~\ref{table:s_dep_comp}) similar to
\objectname[J005252.87-722842.9]{J005252}. The LMC star
(\objectname[J050632.10-714229.8]{J050632}) has a lower
luminosity than \objectname[J005252.87-722842.9]{J005252} while for
the Galactic post-AGB star (\objectname[HD187885]{HD187885}) there is no luminosity estimate
available. Our comparison (see Figure~\ref{comp_s}) shows that, despite the very similar stellar
parameters between the three stars, \objectname[J005252.87-722842.9]{J005252} is clearly not
$s$-process enriched. 

\subsection{Is J005252 depleted?}

\begin{figure}
\begin{center}
\epsscale{1.10}
\plottwo{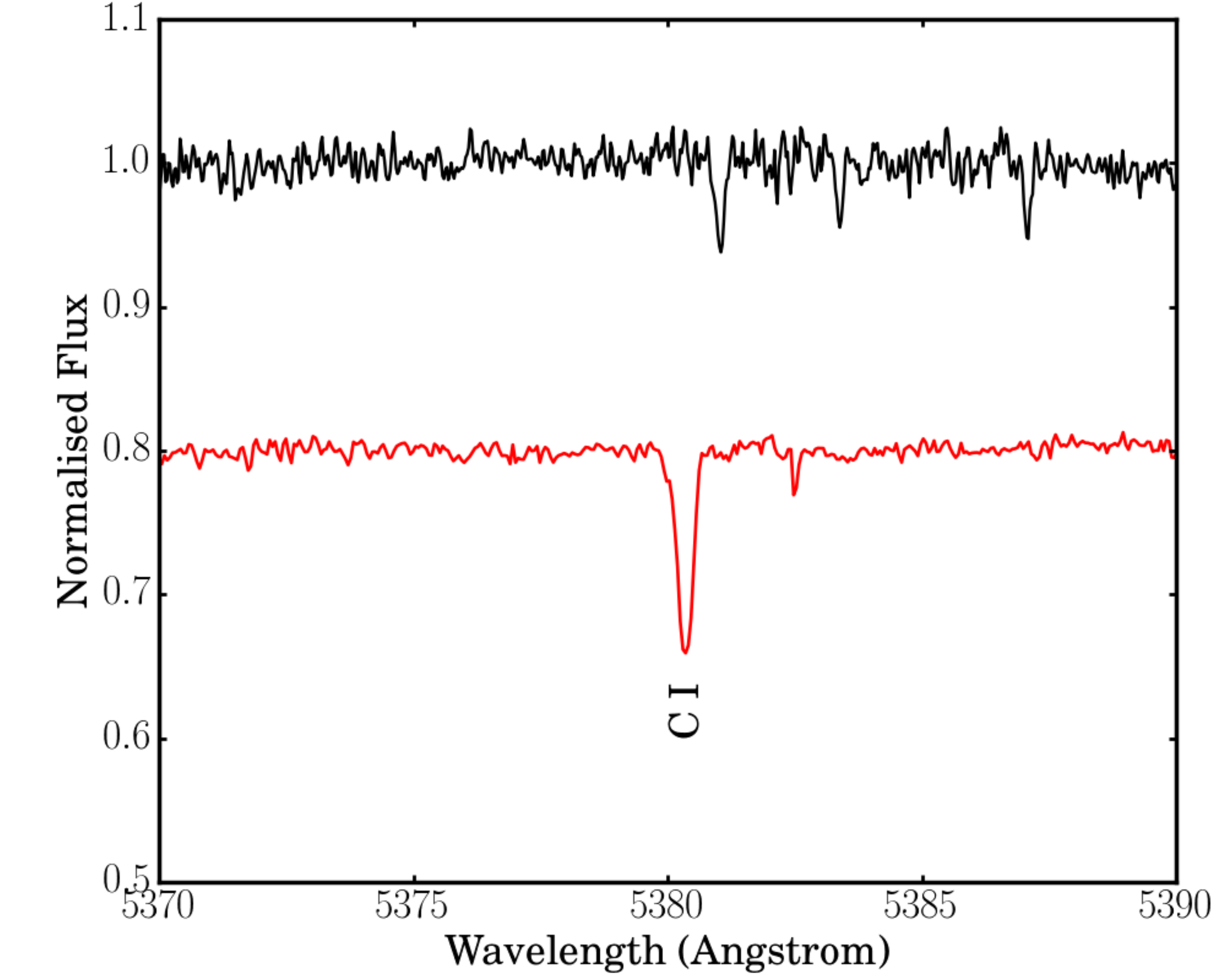}{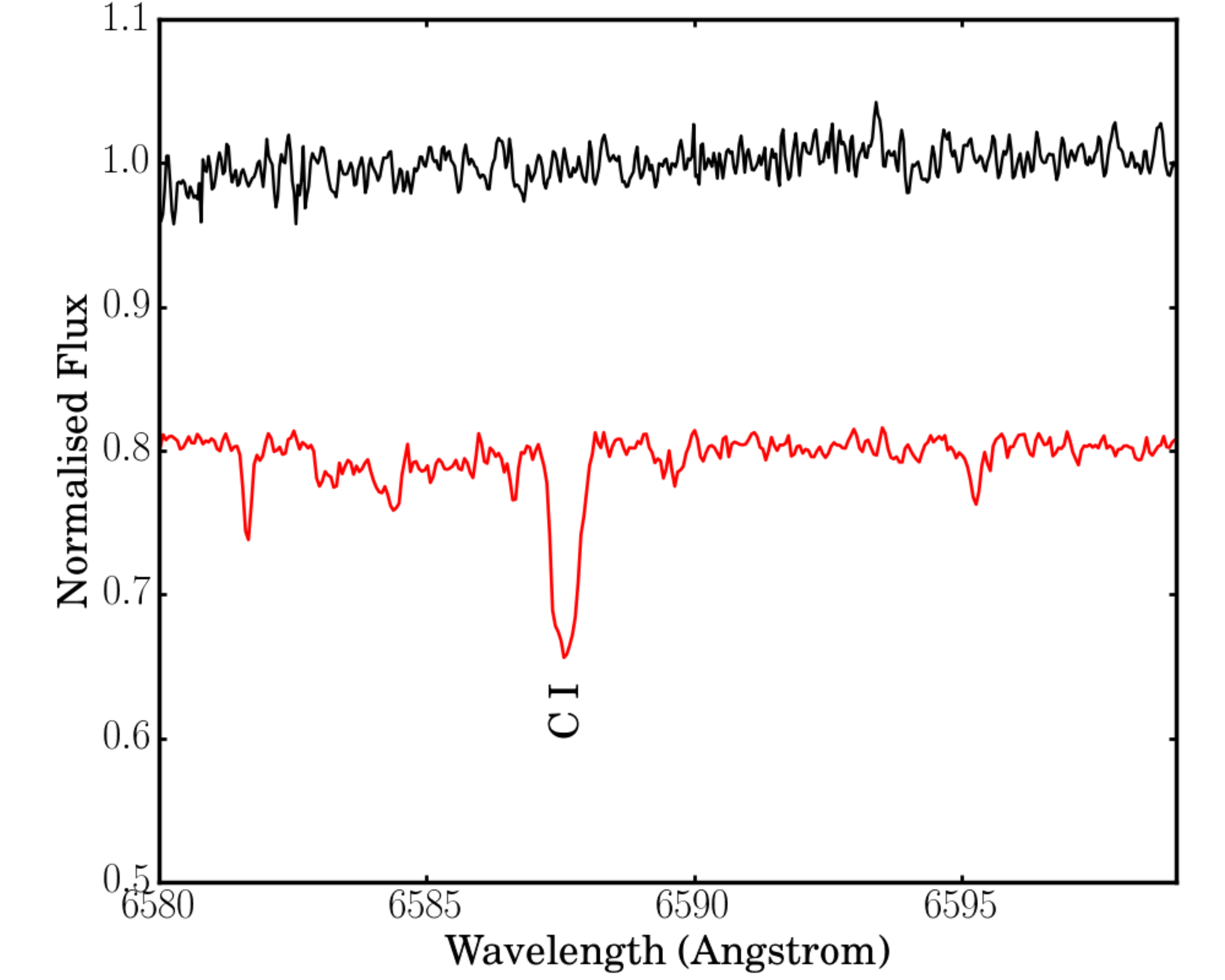}
\caption{Normalised spectra of J005252 (in black) covering the region that
  contains the neutral carbon line (C\,I) line at
  5380.34\,\AA (left panel) and 6587.62 (right panel). In both the plots we show the spectra of a known
  binary post-AGB star BD+394926 (in red), with a depleted
  photospheric chemistry, for comparison.\label{comp_dep}}
\end{center}
\end{figure}

\begin{figure}
\begin{center}
\epsscale{1.10}
\plotone{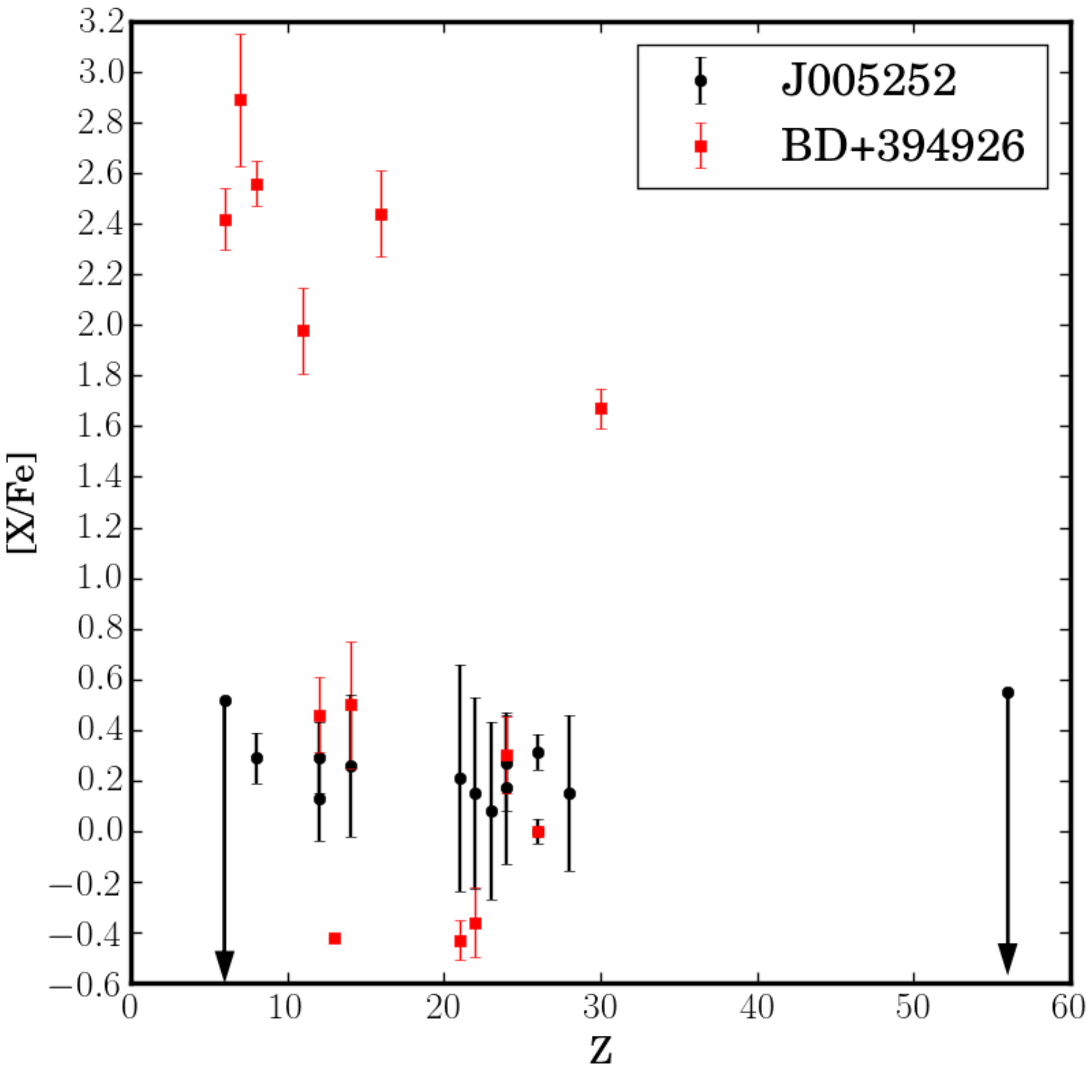}
\caption{Element over iron [X/Fe] ratios for J005252
  (in black) and BD+394926 (in red). Note: For J005252,
the abundances of C\,(Z\,=\,6) and Ba\,(Z\,=\,56) are upper limits and
marked with downward arrows. For J005252, the O\,I and Fe\,I
we plot are the abundances corrected for NLTE effects (see Table~\ref{table:abund_anal})\label{comp_dep_abund}}
\end{center}
\end{figure}

Since \objectname[J005252.87-722842.9]{J005252} does not follow the single star
chemical evolutionary trends we consider the possibility of
photospheric depletion.  We perform a comparison of \objectname[J005252.87-722842.9]{J005252} with a Galactic binary post-AGB star
\objectname[BD+394926]{BD+394926} \citep{rao12, gezer15}. This is the only
known depleted Galactic
post-AGB star with similar stellar parameters (see
Table~\ref{table:s_dep_comp}) to our
object. \objectname[BD+394926]{BD+394926} is very affected by the depletion process which
results in a  \feh\,=\,$-$2.4. 
Post-AGB binaries have a
  [C/Fe]\,$>$\,0 because the difference in condensation temperature
between C and Fe during the depletion process leads to high [C/Fe]. However, \objectname[J005252.87-722842.9]{J005252}
  does not show any traces of carbon in its photospheric
  composition. In Figure~\ref{comp_dep} we compare the
spectral region covering the C\,I lines at 5380.34\,\AA\, and
6587.62\,\AA\, of our object to that of \objectname[BD+394926]{BD+39 4926}. The spectra
of \objectname[BD+394926]{BD+394926} shows the presence of the C\,I
line while our object does not. In Figure~\ref{comp_dep_abund} we show a comparison of the abundance
analysis results ([X/Fe] vs Z) of \objectname[J005252.87-722842.9]{J005252} and
\objectname[BD+394926]{BD+394926}. As expected,
\objectname[BD+394926]{BD+394926} has a high [C/Fe] ratio. Furthermore, \objectname[BD+394926]{BD+394926} shows the presence of volatile elements like Zn
and S, and a depletion of refractory elements that scale with with
Fe. \objectname[J005252.87-722842.9]{J005252} shows no signs of a
depleted photospheric composition (see
Figure~\ref{comp_dep_abund}) and hence is unlikely to be in a binary system.

We conclude that \objectname[J005252.87-722842.9]{J005252} is the
first known object with a photospheric chemistry that does not comply
with the standard theories of stellar evolution and
nucleosynthesis. The well constrained luminosity and stellar
parameters of this object, argue for 
its post-AGB evolutionary nature. It is the most luminous
post-AGB star to be studied in detail to
date. 

Ironically, we note that \objectname[J005252.87-722842.9]{J005252} may
be the first star to be consistent with standard 
AGB evolutionary models. Initially, AGB models for low masses did not show the
production of $s$-process elements because there was no $^{13}$C-pocket formed without
adding some form of partial mixing to the models. With observations making it clear that 
such mixing was occurring, various mechanisms were included in the models to produce such 
a pocket, and the subsequent enhancements in $s$-process elements. Likewise, the
third dredge-up is notoriously difficult to predict quantitatively
\citep{frost96} In many cases some extra-mixing, or overshoot, was added to try to initiate dredge-up at
low masses, to make models consistent with the observations. It is perhaps amusing that
we are now searching for an explanation as to
why \objectname[J005252.87-722842.9]{J005252} matches the original models, and
not those that are constructed to fit the majority of stars.


\section{Galactic Analogues of J005252}
\label{gal_analogues}

\begin{figure}
\begin{center}
\epsscale{1.10}
\plottwo{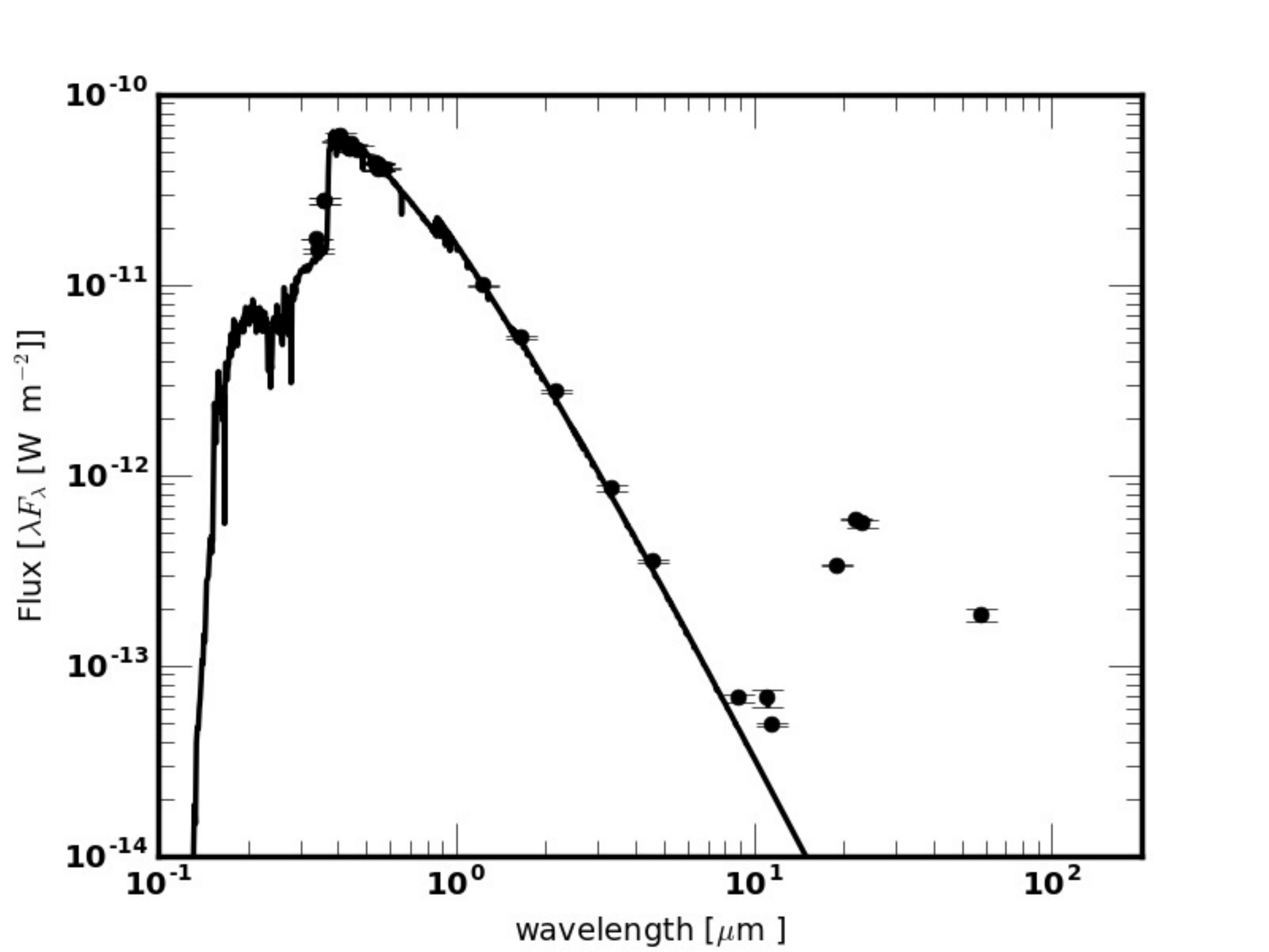}{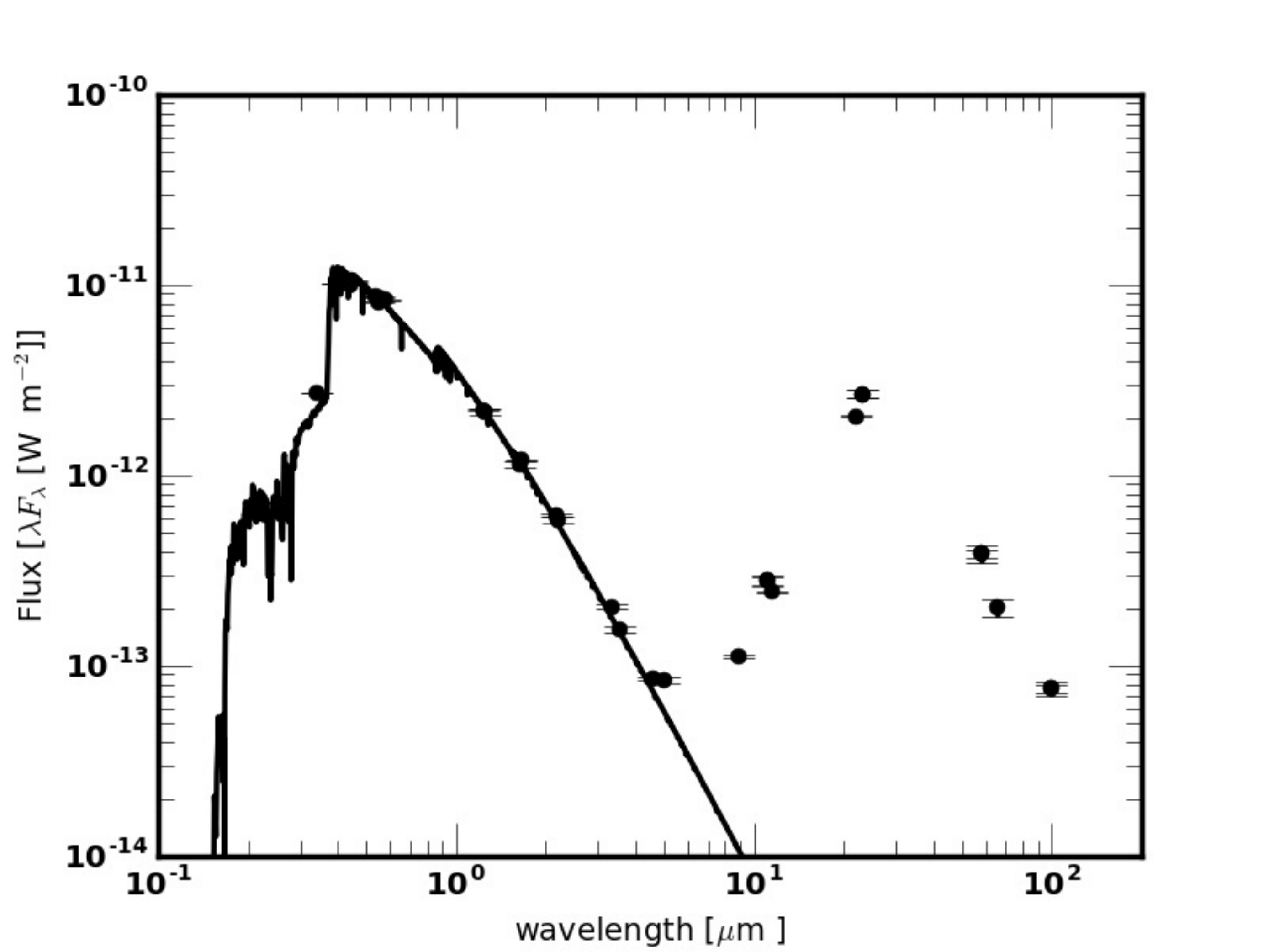}
\caption{Spectral energy distribution of the two likely Galactic
  analogues of J005252: HD133656 (left panel) and
  SAO239854 (right panel). The black symbols
  indicate the broadband photometry corrected for foreground
  extinction. \label{gal_analogues_sed}}
\end{center}
\end{figure}

\begin{figure}
\begin{center}
\epsscale{1.10}
\plottwo{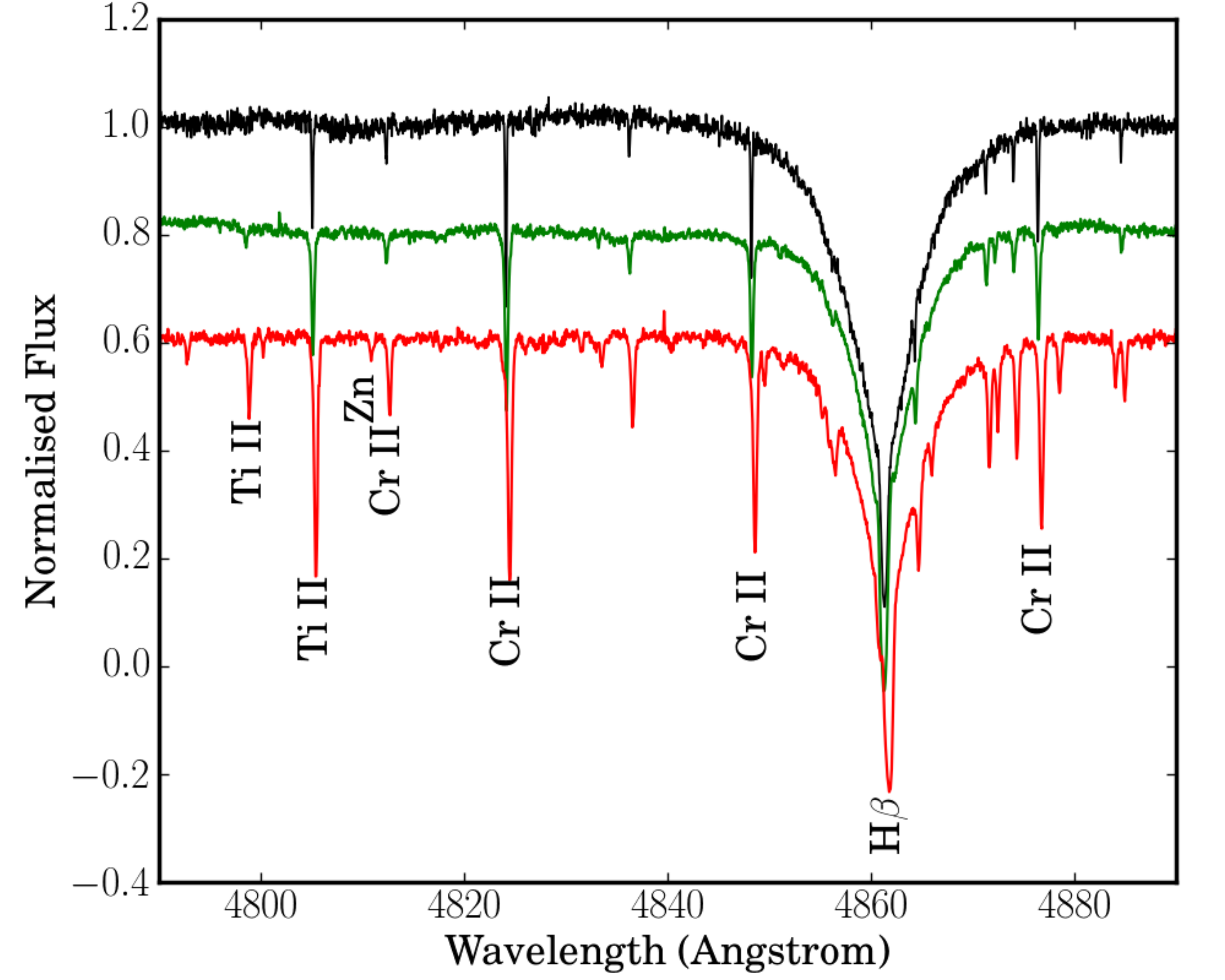}{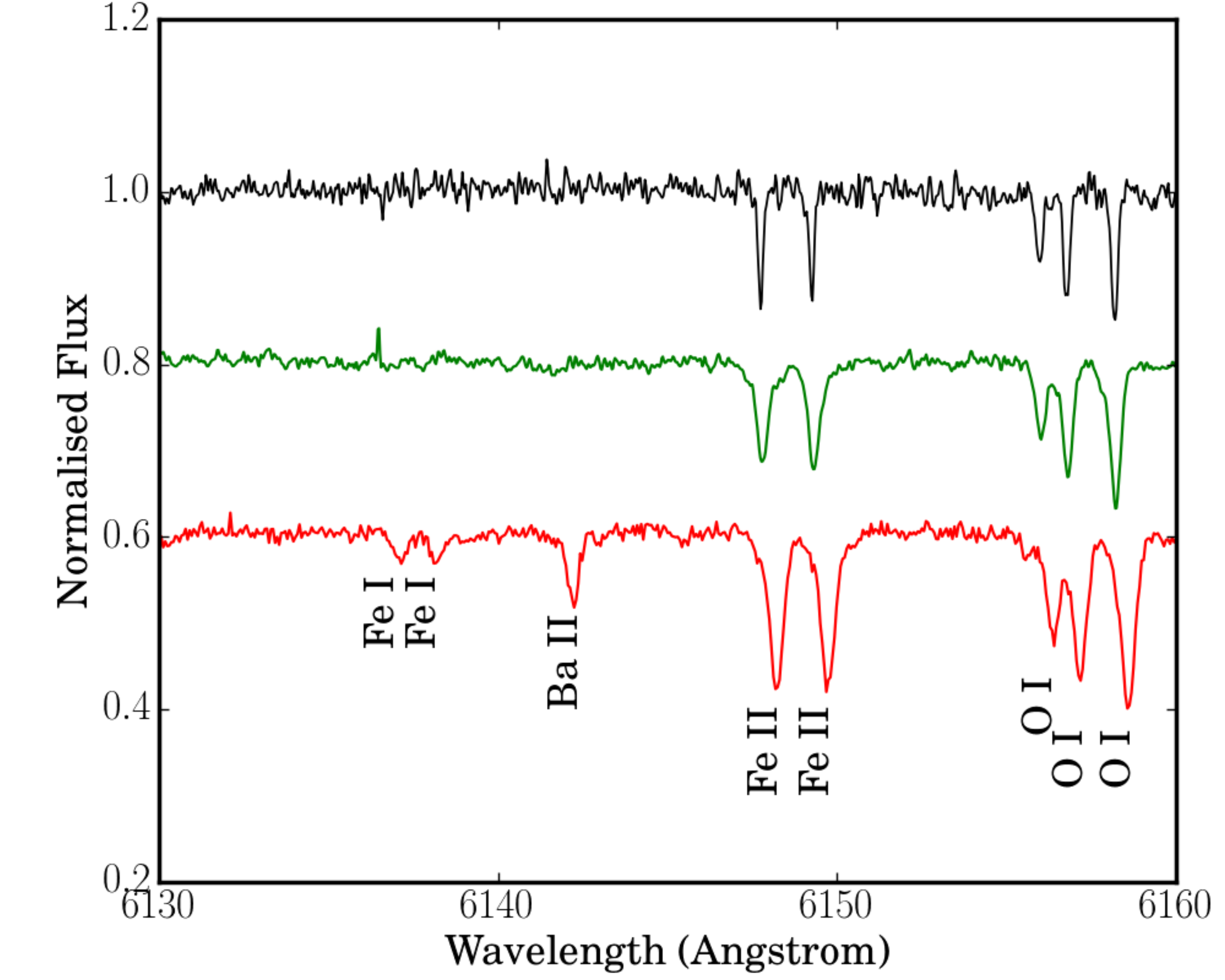}
\caption{Normalised spectral samples of J005252 (in black),
  HD133656 (in green) and SAO239854 (in
  red).\label{gal_analogues_spec}}
\end{center}
\end{figure}

\begin{figure}
\begin{center}
\epsscale{1.0}
\plotone{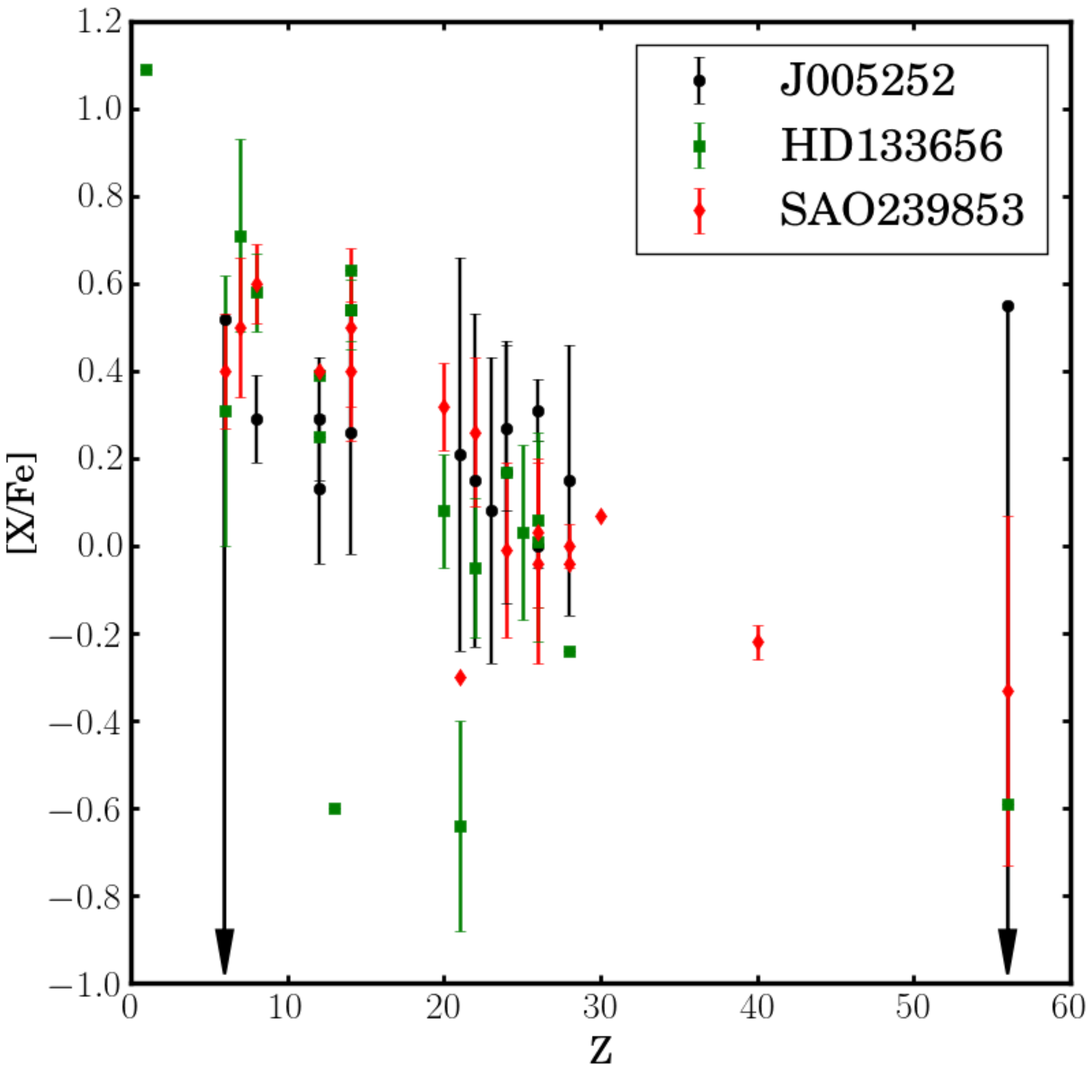}
\caption{Element over hydrogen [X/Fe] ratios of  J005252 (in black),
  HD133656 (in green) and SAO239854 (in red). The error bars respresent the
  total uncertainties $\sigma_{\rm tot}$ in [X/Fe]. Note: For J005252,
the abundances of C\,(Z\,=\,6) and Ba\,(Z\,=\,56) are upper limits and
marked with downward arrows. For J005252, the O\,I and Fe\,I
we plot are the abundances corrected for NLTE effects (see Table~\ref{table:abund_anal}).\label{J005252_gal_XFe}}
\end{center}
\end{figure}

Since the Galactic sample is observationally well studied, we looked
for analogues of \objectname[J005252.87-722842.9]{J005252} amongst the known Galactic post-AGB
stars. We were able to identify thick-disk objects: \objectname{HD133656} and
\objectname{SAO239853}. These two objects have very similar SED
characteristics to \objectname[J005252.87-722842.9]{J005252} (see
Figure~\ref{gal_analogues_sed}) and also similar stellar parameters (see
Table~\ref{table:gal_analogues}). Figure~\ref{gal_analogues_spec} shows the
similarity between the spectra of the three stars. The
photospheric chemistry of \objectname{HD133656} and \objectname{SAO239853} were studied in detail by
\citet{vanwinckel96b} and \citet{vanwinckel97a}, respectively. In
Figure~\ref{J005252_gal_XFe} we plot the [X/Fe] vs atomic mass (Z) for the three objects. We find
that all three objects show similar trends in their photospheric
chemistry (see Figure~\ref{J005252_gal_XFe}), with \objectname[J005252.87-722842.9]{J005252} being the
most extreme case in terms of an unusual chemistry. 
In this section we summarise the abundance
results of the two objects.

\objectname{HD133656} was first identified as a high-latitude metal-deficient star by
\citet{vanwinckel96b}. The photometry, IUE spectrum, H$\gamma$ and H$\beta$ line
profiles all agree very well with an effective temperature of about
8000\,K and a \logg\, of 1.0. The study by \citet{vanwinckel96b}
showed that this star is metal poor (\feh\,$\approx$\,$-$1.00) but with
[C/Fe]\,=\,+0.3, [N/Fe]\,=\,+0.7 and
[O/Fe]\,=\,+0.5. The oxygen abundance follows the oxygen content
of unevolved Galactic stars
of the same metallicity. The light $\alpha$-elements (Mg, Si, S) yield a
[$\alpha$/Fe]\,=\,+0.45. The heavier $\alpha$-element Ca follows the Fe
deficiency, and so do the Fe-peak elements Sc, Ti, Cr and Ni. No
$s$-process overabundances were detected.

\objectname{SAO239853} was first classified as a post-AGB candidate star by
\citet{hrivnak89} on the basis of the IRAS colours. It is a pulsating
variable with a period of 37 days and a total amplitude
of 0.103 magnitudes in $V$-band \citep{bogaert94}. The study by \citet{vanwinckel97a}
showed that this object is metal-deficient with an Fe 
abundance between $-$0.8 and $-$1.0. The
atmosphere is C and N rich, with a [C/Fe]\,=\,+0.4, and
[N/Fe]\,=\,+0.5. The O abundance of [O/Fe]\,=\,+0.6 to +0.7 is about 0.2 dex
higher than expected for an unevolved object with \feh\, between
$-$0.8 to $-$1.0, indicating a slight O enrichment. The star was
identified to have a $s$-process element deficiency based on two small
lines of Zr and two lines of Ba, which yields a [s/Fe] between $-$0.3 and
$-$0.4. 

Amongst the observed sample of post-AGB stars whose chemistry has been
studied in detail these two objects are the best known Galactic analogues of
\objectname[J005252.87-722842.9]{J005252}. The C-enhancement in these
stars is in the order of +0.3 for \objectname{HD133656} and +0.4 for
\objectname{SAO239853}. For metallicities of around \feh\,$=$\,$-$1.0,
stars with thick-disc kinematics have [C/Fe]\,$\approx$\,0.2 \citep{nissen14}. This
suggests that these two stars exhibit a
photospheric composition with no strong C-enhancement. These stars
also do not show a strong $s$-process enrichment or a photospheric
depletion. In these stars, the lack of a nucleosynthetic history that
reflects the third dredge-up was previously associated with the possibility that
their initial mass was too low for active nucleosynthesis enrichment
on the AGB \citep{vanwinckel96a}. Another possibility is that the
objects are indeed of low luminosity and are associated with the
recently identified post-RGB stars \citep[see][and
Section~\ref{intro}]{kamath16}. 

As mentioned in Section~\ref{intro}, the recent first Gaia data
release, (Gaia DR1, Gaia Collaboration et al. 2016), has provided
parallaxes for some Galactic post-AGB stars. Parallaxes are available
for both \objectname{HD133656} and \objectname{SAO239853}. For
\objectname{SAO239853}, the error in parallax (0.34\,mas) is twice the
parallax (0.15\,mas), which results in a large variation in luminosity
(ranging from 1200\,\Lsun to $\>$\,14000\,\Lsun). Therefore, we
are unable to comment whether \objectname{SAO239853} is a post-AGB or
post-RGB object. In the case of \objectname{HD133656}, the
error in parallax (0.34\,mas) is around one third of the parallax
(1.10\,mas). However, this translates into a luminosity that ranges between
800\,\Lsun and 2800\,\Lsun), which implies that the evolutionary
nature of \objectname{HD133656} remains uncertain. With future data
  releases from Gaia, which promise more accurate parallaxes, we
  should soon be able to obtain accurate distances and hence luminosities of these likely
Galactic analogues to confirm their evolutionary nature and therefore their
chemical peculiarities. 

\section{Conclusions}
\label{conclusions}

Atmospheres of post-AGB stars contain all the chemical
enrichment from internal nucleosynthesis that has occurred during the
entire AGB phase and therefore serve as excellent probes to understand
stellar evolution and nucleosynthesis. 
Galactic post-AGB stars have been found to be
chemically much more diverse than anticipated. Some objects are the
most $s$-process enriched objects known to date while others are not
enriched at all. However, the poorly known distances (and hence
luminosities and masses) of the Galactic post-AGB sample hamper the
interpretation of their abundances in the broader theoretical context
of stellar  evolution. Due to their known distances, MC 
post-AGB stars provide excellent constraints and offer unprecedented tests for
AGB theoretical evolution and nucleosynthesis models of low- and intermediate-mass stars. Our recent search for MC
post-AGB stars \citep[][in the SMC and LMC, respectively]{kamath14,kamath15} yielded a spectroscopically verified catalogue of
optically visible post-AGB stars. We are currently doing follow-up
high-resolution detailed chemical abundance studies of a few of these
objects, one of which is \objectname[J005252.87-722842.9]{J005252}. 

In this paper, 
we argue for the likely post-AGB nature of 
\objectname[J005252.87-722842.9]{J005252} and present a detailed
photospheric study of this object which revealed its 
intriguing photospheric composition. This luminous (8200\,$\pm$\,700\,\Lsun), metal-poor
(\feh\,=\,$-$\,1.18\,$\pm$\,0.10) star with a shell-type SED, provides the first observational
evidence for a star that fails the third dredge-up. This object with
\teff\,=\,8250\,$\pm$\,250\,K, \logg\,=\,1.0\,$\pm$\,0.25\,dex, and
\mv\,=2.0\,$\pm$\,0.25\,kms$^{\rm -1}$, shows no traces of
carbon-enhancement, indicating that this low-metallicity, low initial mass
(1.5 to 2.0\Msun) object has not undergone any third dredge-up episode
during its entire AGB phase. The oxygen is 
above solar with [O/Fe]\,=\,0.29\,$\pm$\,0.1, after taking into
account NLTE corrections. Furthermore, \objectname[J005252.87-722842.9]{J005252} shows neither signs of an 
$s$-process enrichment, which is characteristic of single post-AGB stars, nor a depleted
photospheric chemistry which is characteristic of most binary post-AGB stars. 

Assuming that \objectname[J005252.87-722842.9]{J005252} is a single
star, stellar evolution and nucleosynthesis models \citep{lattanzio86,lattanzio89,karakas10a,cristallo11,fishlock14} that are
appropriate for \objectname[J005252.87-722842.9]{J005252} (based on
its luminosity, initial mass and metallicity) predict this object to
undergo third dredge-up and become C-rich. Synthetic
models of TP-AGB stars that are calculated to fit the SMC CSLF also
predict a star 
with the luminosity and metallicity of
\objectname[J005252.87-722842.9]{J005252} to be C-rich. 

We consider the possibilities of nucleosynthetic process, such as
HBB \citep{boothroyd93,lattanzio92,lattanzio96} that can prevent the formation of
high-luminosity carbon stars. However, at the metallicity of
\objectname[J005252.87-722842.9]{J005252} ( Z\,=\,0.001), HBB is only active in stars
with $M$\,$\gtrsim$\,2.5\,\Msun \citep{ventura14}. Since \objectname[J005252.87-722842.9]{J005252} is star with
$M$\,$\approx$\,1.5\,$-$\,2\Msun and Z\,=\,0.001, it is likely that
the temperature at the base of the convective envelope during the
interpulse period is not high enough to activate HBB. 

Therefore, the only way to explain the intriguing chemistry of
\objectname[J005252.87-722842.9]{J005252} is via some mechanism that
results in an AGB life without dredge-up episodes.  Alternatively,
this object could be a product of a merger scenario wherein it is not
easy to trace the merger process and therefore understand the
nucleosynthetic history of the end-product. However, the spectral
lines of \objectname[J005252.87-722842.9]{J005252} are resolved and do
not shows signs of strong rotation, which is normally a characteristic feature of mergers.

In our attempt to find
Galactic analogues, we found two stars: \objectname{SAO239853} and \objectname{HD133656}, which are
likely to also have failed the third dredge-up. However, their
uncertain luminosities brings in the possibility of them being post-RGB stars and therefore restricts us from
confirming the possibility of a failed TDU in these stars.

\objectname[J005252.87-722842.9]{J005252} is the first detected object of its
kind. This object along with its two likely Galactic analogues opens up a possibility for a new evolutionary channel or
unknown mass-loss and/or nucleosynthetic processes that govern the AGB phase. One of the useful probes of better understanding the chemistry of
\objectname[J005252.87-722842.9]{J005252} is the N abundance as it is
affected via the CNO cycle. Our current
spectral region does not cover usable N lines. We aim to obtain
spectra covering the N lines to be able to determine if whether the N
abundance can give us hints for solving the nucleosynthetic mystery of
\objectname[J005252.87-722842.9]{J005252}. 




\section{Acknowledgments}
DK and HVW acknowledge the support of the KU Leuven contract
GOA/13/012. DK acknowledges the support of the FWO grant G.OB86.13. 
PRW has received support from the Australian Research
Council Discovery Project DP120103337.



{\it Facilities:} \facility{This study is based on observations collected with the Very Large Telescope
  at the ESO Paranal Observatory (Chili) of programme number 092.D-0485}



\bibliography{mnemonic,devlib}

\clearpage

\clearpage

\clearpage

\begin{table}
\begin{center}
\caption{Target details and overview of observations.\label{table:fields}}
\begin{tabular}{cccccccc}
\tableline\tableline
Star & RA & DEC & Exp.time\,(s) & Exp time\,(s) &
$V$ & $L_{\rm obs}$/\Lsun & RV \\
       & (J2000) & (J2000) & Blue Arm & Red Arm & mag & & kms$^{-1}$\\
\tableline
J005252 & 00:52:52.87 & -72:28:42.9 & 3\,$\times$\,1500 &
3\,$\times$\,1500 & 14.031 & 8000 & 149\,$\pm$\,1\\
\tableline
\end{tabular}
\tablecomments{'Exp.Time' represents the exposure time. 
The Red arm is the combination of the lower and upper part of the
mosaic CCD chip. $V$ represents the $V$-band magnitude, $L_{\rm
  obs}$/\Lsun\, is the observed luminosity corrected for foreground extinction, RV represents the
heliocentric radial velocity (see text for
details).}
\end{center}
\end{table}

\clearpage

\begin{table}
\caption{Spectroscopically determined atmospheric parameters of
  J005252.\label{table:atmos_param}}
\begin{center}
\begin{tabular}{lcc} \hline\hline
Stellar Parameters &  J05252$_{\rm UVES}$  &  J005252$_{\rm AAO}$   \\   
\hline
\teff\,(K) & 8250\,$\pm$\,250  & 7600\,$\pm$\,500 \\
\logg\,(dex)    & 1.0\,$\pm$\,0.25 & 1.4\,$\pm$\,0.50 \\
\mv\,(kms$^{-1}$) & 2.0\,$\pm$\,0.25   & - \\ 
\feIh  &  $-$1.05\,$\pm$\,0.34 & - \\
\feIh$_{\rm NLTE}$  &  $-$0.87\,$\pm$\,0.34 & - \\
\feIIh &  $-$1.18\,$\pm$\,0.10 & - \\
\feh &  $-$1.18\,$\pm$\,0.10 &  $-$2.00\,$\pm$\,0.5\\
N$_{\textrm{FeI}}$  & 34  &  - \\
N$_{\textrm{FeII}}$ & 31  &   - \\
E($B$-$V$) & 0.02\,$\pm$\,0.02 & 0.02\,$\pm$\,0.02 \\
$L_{\rm ph}$/\Lsun & 8200\,$\pm$\,700  & 7500 \\
$M$/\Msun & 0.63\,$\pm$\,0.02 & 0.62 \\
\hline
\end{tabular}
\tablecomments{J05252$_{\rm UVES}$ represents the spectroscopically determined atmospheric parameters of
  J005252 obtained using the UVES spectra presented in this study.  J005252$_{\rm AAO}$ represents the spectroscopically determined atmospheric parameters of
  J005252 obtained using the low-res optical AAOmega spectra presented in
  \citep{kamath14}. We take the estimated \feIIh\, value as the
  metallcity (\feh) of the star. \feIh$_{\rm NLTE}$ represents the value of \feIh\, after
  taking into account the NLTE effects for Fe\,I. We note that the
  NLTE effects for Fe\,II are negligible (see Section~\ref{atmos_param} for full
  details). The errors for \feh\, include line to line scatter and model uncertainty. N$_{\textrm{FeI}}$ and N$_{\textrm{FeII}}$ show the number of lines used
for \feI\, and \feII\, respectively. Note: Assumed solar [Fe/H]\,=\,7.50\,dex
\citep{asplund09}. $L_{\rm ph}$/\Lsun represents the photospheric luminosity of J005252
and  $M$/\Msun is the derived mass of \objectname[J005252.87-722842.9]{J005252} (see
Section~\ref{lum+ini_mass} for details).}
\end{center}
\end{table}

\clearpage

\begin{table}
\begin{center}
\caption{\label{table:linelist}The linelist of J005252.87-722842.9.}
\setlength{\tabcolsep}{0.15cm}
\begin{tabular}{lccccc } \hline\hline
Ion & $\lambda$ & EP & $\log$\,gf & EW & Abundance\\
\hline
\hline
8.0 & 6155.971 & 10.740 & -0.670 & 19.40 & 8.258\\
8.0 & 6156.778 & 10.740 & -0.450 & 26.80 & 8.220\\
8.0 & 6158.187 & 10.740 & -0.310 & 33.70 & 8.219\\
12.0 & 5167.320 & 2.700 & -0.750 & 45.80 & 6.596\\
12.0 & 5528.420 & 4.350 & -0.470 & 9.10 & 6.577\\
\hline
\end{tabular}
\tablenotes{Ion denotes the atomic number of the respective element
  along with the ionisation stage. $\lambda$ represents the rest frame
  wavelength (in $\AA$), EP is the excitation potential in eV,
  $\log$\,gf is the oscillator strength, EW is the equivalent width in
  m$\AA$. The full table is available as online supporting
  information.}
\end{center}
\end{table}
\clearpage

\clearpage

\begin{table}
\begin{center}
\caption{\label{table:abund_anal} Abundance results of J005252.}
\setlength{\tabcolsep}{0.15cm}
\begin{tabular}{lccccccccc } \hline\hline
Ion   & Z & N & $\log$\,$\epsilon_{\rm o}$  & $\log\,\epsilon$  &
$\sigma$$_{\rm l2l}$ & [X/H] & $\sigma_{\rm tot}$[X/H] & [X/Fe]
& $\sigma_{\rm tot}$[X/Fe]\\
\hline
O\,I & 8 & 3 & 8.69 & 8.23 & 0.06 & $-$0.47 & 0.06 & 0.71 & 0.1\\
O\,I$_{\rm NLTE}$ & 8 & 3 & 8.69 & 8.23 & 0.06 & $-$0.89 & 0.06 & 0.29 & 0.1\\
Mg\,I & 12 & 6 & 7.6 & 6.55 & 0.19 & $-$1.05 & 0.30 & 0.13 & 0.17\\
Mg\,II & 12 & 3 & 7.6 & 6.71 & 0.12 & $-$0.89 & 0.10 & 0.29 & 0.14\\
Si\,II & 14 & 5 & 7.51 & 6.60 & 0.34 & $-$0.92 & 0.29 & 0.26 & 0.28\\
Sc\,II & 21 & 1 & 3.15 & 2.18 & 0.2 & $-$0.97 & 0.44 & 0.21 & 0.45\\
Ti\,II & 22 & 23 & 4.95 & 3.92 & 0.4 & $-$1.03 & 0.39 & 0.15 & 0.38\\
V\,II & 23 & 5 & 3.93 & 2.83 & 0.1 & $-$1.1 & 0.34 & 0.08 & 0.35\\
Cr\,I & 24 & 3 & 5.64 & 4.73 & 0.19 & $-$0.91 & 0.32 & 0.27 & 0.19\\
Cr\,II & 24 & 15 & 5.64 & 4.63 & 0.09 & $-$1.01 & 0.3 & 0.17 & 0.3\\
Fe\,I & 26 & 34 & 7.5 & 6.39 & 0.11 & $-$1.05 & 0.34 & 0.13 & 0.07\\
Fe\,I$_{\rm NLTE}$ & 26 & 34 & 7.5 & 6.39 & 0.11 & $-$0.87 & 0.31 & 0 & 0.07\\
Fe\,II & 26 & 39 & 7.5 & 6.48 & 0.08 & $-$1.18 & 0.10 & 0 & 0.05\\
Ni\,II & 28 & 2 & 6.22 & 5.19 & 0.18 & $-$1.03 & 0.32 & 0.15 & 0.31\\
\hline
C\,I & 6 & & 8.43 & 7.69 & & $<$\,$-$0.66 & & $<$\,0.50 & \\
Ba\,II & 56 & & 2.18 & 1.55 & & $<$\,$-$0.63 & & $<$\,0.55 & \\
\hline
\end{tabular}
\tablenotes{The first column gives the ions that
were detected and studied. The second column gives their corresponding
atomic number. N is the number of lines used for each ion. $\log$\,$\epsilon_{\rm o}$ are the solar
abundances of the elements from \citet{asplund09}, $\log\,\epsilon$ is
the determined abundance,
$\sigma$$_{\rm l2l}$ is the line-to-line scatter, [X/Fe] is the
element-over-iron (FeII) ratio, [X/H] is the element-over-hydrogen ratio,
and $\sigma_{\rm tot}$[X/H] and $\sigma_{\rm tot}$[X/Fe]  is the
total uncertainty on [X/H] and [X/Fe], respectively. We impose a $\sigma$$_{\rm l2l}$ of 0.20 dex for all 
ions for  which only one line is available for the abundance
determination. \feIh$_{\rm
  NLTE}$ and O\,I$_{\rm
  NLTE}$ represents the values after
  taking into account the NLTE effects for Fe\,I. We note that the
  NLTE effects for Fe\,II are negligible and unknown for other
  elements (see Section~\ref{atmos_param} for full
  details). We note that the
abundances of C\,(Z\,=\,6) and Ba\,(Z\,=\,56) are upper limits. See text for further details.}
\end{center}
\end{table}

\clearpage

\clearpage

\clearpage

\clearpage

\clearpage

\begin{table}
\caption{Atmospheric parameters of
  J005252, J050632, HD187885 and BD+394926.\label{table:s_dep_comp}}
\begin{center}
\begin{tabular}{lcccc} \hline\hline
Stellar Parameters &  J052522&  J050632 &  HD187885 & BD+394926 \\   
\hline
Host galaxy & SMC & LMC & Milky Way & Milky Way \\
\teff\,(K) & 8250\,$\pm$\,250  & 6750\,$\pm$\,125 & 6750\,$\pm$\,125 & 7750\,$\pm$\,250\\
\logg\,(dex)    & 1.00\,$\pm$\,0.25 &  1.0\,$\pm$\,0.25 &  1.5\,$\pm$\,0.25  &  1.0\,$\pm$\,0.5\\
\mv\,(kms$^{-1}$) & 2.0\,$\pm$\,0.25   &  3.0\,$\pm$\,0.25 &  5.0\,$\pm$\,0.25 &  3.0\,$\pm$\,0.5\\ 
\feh &  $-$1.18\,$\pm$\,0.10 &  $-$1.22\,$\pm$\,0.16 & $-$0.5\,$\pm$\,0.20 &  $-$2.37\,$\pm$\,0.20\\
E($B$-$V$) & 0.02\,$\pm$\,0.02 & 0.05\,$\pm$\,0.06 & - & - \\
$L_{\rm ph}$/\Lsun & 8200\,$\pm$\,700,  & 5400\,$\pm$\,700 & - & - \\
\hline
\end{tabular}
\tablecomments{ Columns 1,2,3 and 4 list the spectroscopically determined atmospheric parameters of
  J005252, J050632, HD187885 and BD+394926 obtained using the UVES spectra presented in this study, 
from \citet{vanaarle13}, from \citet{vanwinckel96a} and from \citet{rao12}, respectively. We note that for the Galactic objects  
HD187885 and BD+394926 there are no luminosity and E($B$-$V$) measurements available.}
\end{center}
\end{table}

\clearpage

\clearpage

\clearpage

\clearpage

\clearpage

\begin{table}
\caption{Atmospheric parameters of
  J005252 and its likely Galactic analogues: HD133656 and SAO239853.\label{table:gal_analogues}}
\begin{center}
\begin{tabular}{lcccc} \hline\hline
Stellar Parameters &  J005252 & HD133656 & SAO239853  \\   
\hline
Host galaxy & SMC & Milky Way & Milky Way \\
\teff\,(K) & 8250\,$\pm$\,250  & 8000\,$\pm$\,125 & 7500\,$\pm$\,125 \\
\logg\,(dex)    & 1.0\,$\pm$\,0.25 &  1.0\,$\pm$\,0.25 &  1.0\,$\pm$\,0.25 \\
\mv\,(kms$^{-1}$) & 2.0\,$\pm$\,0.25  &  6.0\,$\pm$\,0.25 &  5.0\,$\pm$\,0.25 \\ 
\feh &  -1.18\,$\pm$\,0.10 & -1.0\,$\pm$\,0.16 & -0.8 \,$\pm$\,0.20\\
E($B$-$V$) & 0.02\,$\pm$\,0.02 & - & - \\
$L_{\rm ph}$/\Lsun & 8200\,$\pm$\,700 & - & - \\
\hline
\end{tabular}
\tablecomments{ Columns 1,2 and 3 list the spectroscopically determined atmospheric parameters of
  J005252, HD133656 and SAO239853 obtained using the UVES spectra presented in this study, 
from \citet{vanwinckel96b}, from \citet{vanwinckel97a}, respectively. We note that for the Galactic objects HD133656 and SAO239853, 
there are no luminosity and E($B$-$V$) measurements available.}
\end{center}
\end{table}

\clearpage

\clearpage

\clearpage

\end{document}